\documentclass[lettersize,journal]{IEEEtran}
\usepackage{amsmath,amssymb,amsfonts}
\usepackage{graphicx}
\usepackage{mathrsfs}
\usepackage{array}
\usepackage[caption=false]{subfig}
\usepackage{textcomp}
\usepackage{stfloats}
\usepackage{url}
\usepackage{verbatim}
\usepackage{graphicx,color}
\usepackage[table]{xcolor}
\usepackage{multirow}
\usepackage{booktabs}
\usepackage[most]{tcolorbox}
\usepackage{hyperref}

\definecolor{colortDCF}{RGB}{0, 103, 181}
\definecolor{colorEER}{RGB}{210, 72, 24}
\definecolor{coloredit}{rgb}{0,0,0}

\usepackage[
backend=biber,
style=ieee,
maxbibnames=3,
maxcitenames=3,
doi=false,isbn=false,url=false,eprint=false
]{biblatex} 

\defbibheading{bibliography}[\refname]{}

\begin{document}

\title{ASVspoof~5: Evaluation of Spoofing, Deepfake, \\ and Adversarial Attack Detection \\ Using Crowdsourced Speech}
\author{Xin Wang, \emph{Member, IEEE}, Héctor Delgado, Nicholas Evans, \emph{Member, IEEE}, Xuechen Liu, \emph{Member, IEEE}, Tomi Kinnunen, \emph{Member, IEEE}, Hemlata Tak, \emph{Member, IEEE}, Kong Aik Lee, \emph{Senior Member, IEEE}, Ivan Kukanov, Md Sahidullah, \emph{Member, IEEE}, Massimiliano Todisco, \emph{Member, IEEE}, Junichi Yamagishi, \emph{Senior Member, IEEE} 
\thanks{Xin Wang, Xuechen Liu, and Junichi Yamagishi are with National Institute of Informatics, Tokyo 101-8430, Japan (e-mail: wangxin@nii.ac.jp, xuecliu@nii.ac.jp, jyamagis@nii.ac.jp). Xin Wang is the corresponding author.}
\thanks{Héctor Delgado is with Microsoft, P.º Club Deportivo, 1, Edificio 1, 28223 Pozuelo de Alarcón, Madrid, Spain (e-mail: hector.delgado@microsoft.com).}
\thanks{Nicholas Evans and Massimiliano Todisco are with Digital Security Department, EURECOM (Campus SophiaTech), 06410 Biot, France (e-mail: todisco@eurecom.fr, evans@eurecom.fr).}
\thanks{Tomi Kinnunen is with School of Computing, University of Eastern Finland, FI-80101, Joensuu, Finland (e-mail: tomi.kinnunen@uef.fi).}
\thanks{Hemlata Tak is with Pindrop, 1115 Howell Mill Rd NW \#700, 30318, Atlanta GA, USA (e-mail: Hemlata.Tak@pindrop.com)}
\thanks{Kong Aik Lee is with the Department of Electrical and Electronic Engineering and the Research Centre for Data Science \& Artificial Intelligence, The Hong Kong Polytechnic University, Kowloon, Hong Kong (email: kong-aik.lee@polyu.edu.hk)}
\thanks{Ivan Kukanov is with KLASS Engineering and Solutions, 30A Kallang Pl, \#11-03, 339213 Singapore (email: Ivan@kukanov.com)}
\thanks{Md Sahidullah is with Institute for Advancing Intelligence, TCG CREST, 700091, Kolkata, India (email: sahidullahmd@gmail.com)}
}

\markboth{Journal of \LaTeX\ Class Files,~Vol.~14, No.~8, August~2021}%
{Shell \MakeLowercase{\textit{et al.}}: A Sample Article Using IEEEtran.cls for IEEE Journals}

\maketitle

\begin{abstract}
ASVspoof~5 is the fifth edition in a series of challenges which promote the study of speech spoofing and deepfake detection solutions. 
A significant change from previous challenge editions is a new crowdsourced database collected from a substantially greater number of speakers under diverse recording conditions, and a mix of cutting-edge and legacy generative speech technology. 
With the new database described elsewhere, we provide in this paper an overview of the ASVspoof~5 challenge results for the submissions of 53 participating teams. 
While many solutions perform well, performance degrades under adversarial attacks and the application of neural encoding/compression schemes. 
Together with a review of post-challenge results, we also report a study of calibration in addition to other principal challenges and outline a road-map for the future of ASVspoof.
\end{abstract}


\section{Introduction}

\IEEEPARstart{B}{iometric} systems are known to be vulnerable to \emph{spoofing attacks}, also referred to as presentation attacks~\cite{ISOpresentationAtack}, whereby an adversary attempts to masquerade as another individual through the presentation of artificially generated or manipulated biometric data.  
Automatic speaker verification (ASV) systems are no exception~\cite{wu2015spoofing}. 
The threat posed by speech spoofing attacks, be it to ASV systems or human listeners, has grown with the rapid evolution in deep neural network (DNN)-based, zero-shot voice cloning technology which allows an adversary to forge speech recordings in another speaker's voice using only a few seconds of speech collected from the victim~\cite{casanova2022yourtts,chen_neural_2025}. 
The plethora of publicly available text-to-speech (TTS) and voice conversion (VC) toolkits or APIs~\cite{hayashi2020espnet, Eren_Coqui_TTS_2021, lux2021toolkit, tan_neural_2023, Harper_NeMo_a_toolkit, Elevenlabs} mean that spoofing attacks can even be generated without any specialised expertise. 
Furthermore, the perceived quality of synthetic or converted speech generated with state-of-the-art techniques has reached a level where human listeners can no longer distinguish between spoofed\footnote{
Synthetic data that do not aim to deceive an ASV system but forge utterances in target speakers' voices are referred to as deepfake~\cite{liuASVspoof2021Spoofed2023}. For simplicity, we use the term `spoofed' throughout the paper and distinguish between `spoofed' and `deepfake' only when necessary.
} and bona fide speech recordings~\cite{shen2018natural}.

While others have emerged, e.g., the Audio Deep synthesis Detection (ADD)~\cite{add2022,add2023} and Synthetic Audio Forensics Evaluation (SAFE)~\cite{kirill_safe_2025} challenges, and the recent Interspeech 2025 special session on source tracing 2025~\cite{UsingMLAADforSourceTracing}, 
the ASVspoof initiative and challenge series were founded following the first Interspeech special session on the topic in 2013 to foster the development of  countermeasures (CMs) to protect ASV systems and human listeners from spoofing attacks. 
The first challenge edition held in 2015~\cite{Wu-ASVspoof2015} focused on the development of CMs for the detection of TTS and VC attacks. ASVspoof 2019 was the first to consider the detection of DNN-based spoofing attacks, i.e.\ deepfakes, generated using, e.g., WaveNet~\cite{oord2016wavenet} and Tacotron~\cite{wang_tacotron_2017}.  
ASVspoof 2021 featured more diverse spoof/deepfake attacks and data collected from the 2020 Voice Conversion Challenge~\cite{Yi2020} in addition to transmission and compression variability. 
Alongside a broadening scope of attacks,  ASVspoof has also promoted advances in spoofing-robust ASV and the joint evaluation of combined spoofing and speaker detection solutions.

The latest ASVspoof~5 challenge adopts a different source database to all previous editions. To support the study of spoofing-robust automatic speaker verification, it contains data collected from almost two thousand speakers, an order of magnitude increase compared to previous editions. 
To support the development of more robust solutions, \textcolor{coloredit}{the data exhibit substantially greater variability in recording environments.}
To keep pace with developments in generative speech technology, spoofed data, collected in collaboration with an international team of data contributors, are generated with a diverse blend of the very latest TTS and VC technology, in addition to legacy algorithms. Bona fide and spoofed data are processed with a number of different encoding schemes, including DNN-based codecs, while adversarial attacks are included for the first time.

A description of the ASVspoof~5 database is available in~\cite{wang_asvspoof5database}.
The focus in this paper is upon results, calibration and other principal challenges.  
An outline of the evaluation setup is illustrated in Table~\ref{tab:track-summary}.
There are two tasks, namely the design of stand-alone CMs (spoof/deepfake speech detectors) and of spoofing-robust ASV systems.
For each task there are two evaluation conditions. 
A closed condition was defined to protect evaluation integrity, whereby competing solutions can be compared under otherwise identical data conditions. 
Data used for training, development and evaluation was restricted to a specific, closed set. The use of any other speech data was prohibited.
A second, open condition was also adopted to explore performance when massive collections of shared, public speech data are used by detection system designers and adversaries alike.\footnote{\textcolor{coloredit}{The ASVspoof~5 challenge focuses exclusively on post-sensor attacks launched in a logical access scenario. Accordingly, we do not consider the detection of sensor-level attacks such as replayed speech. Their consideration requires an independent data curation pipeline and remains within scope for future challenge editions (\S~\ref{sec:future}).}}

In extending substantially upon preliminary results presented in~\cite{wang2024asvspoof5}, we present an analysis of principal techniques common to the top-performing systems for each track and condition, and influential data factors that impact system performance (\S~\ref{sec:results}). \textcolor{coloredit}{The impacts of adversarial attacks and encoding, alongside observations across the two evaluation conditions, have not been presented in previous publications.}
Also presented is an analysis of evaluation results using calibration-aware metrics, a first within the ASVspoof challenge series (\S~\ref{sec:dis:calibration_cm}). 
\textcolor{coloredit}{We also include results for a new cross-database evaluation which is included to assess the generalization performance of top submissions (\S~\ref{sec:dis:cross_db_eval}).}

The new insights presented in this article will be of interest to readers working in speech spoof/deepfake detection, 
hence some familiarity with the topic is assumed. 
We nonetheless provide an outline of the ASVspoof~5 challenge (\S~\ref{sec:outline}), before describing both evaluation metrics (\S~\ref{sec:metrics}) and results (\S~\ref{sec:results}) with details of top-performing systems. 
We conclude with a reflection upon the limitations of, and key lessons learned from the ASVspoof~5 challenge, with a discussion of ideas and directions for future research.

\begin{table*}[t!]
   \centering
    \caption{Summary of the detection scenarios, evaluation metrics and system requirements for the ASVspoof~5 challenge Track 1 and Track 2.
    For `classes', star (\texttt{*}) indicates the `positive' class which should be associated with higher detection scores. 
    Participants submit the required scores, and the binary decisions of \texttt{ACCEPT} or \texttt{REJECT} are performed by the organisers.
    }
    \vspace{-2mm}
    \label{tab:track-summary}
    {
    \begin{tabular}{rll}
        \toprule
        & \textbf{Track 1} & \textbf{Track 2} \\
        \midrule
        {Task} & Stand-alone spoof/deepfake detection & Spoofing-robust ASV \\
        {Scenario} & Generic & Telephony or VoIP\\ 
        Classes & \texttt{bonafide*, spoof} & \texttt{target*, nontarget, spoof} \\
        Decisions & \texttt{ACCEPT, REJECT} & \texttt{ACCEPT, REJECT}\\
        Metrics & minDCF ({primary}), actDCF, $C_\text{llr}$ \cite{Brummer2006-application-independent}, EER & min a-DCF \cite{shim2024adcf} ({primary}), min t-DCF \cite{Kinnunen2020-tDCF}, t-EER \cite{Kinnunen2024-tEER}\\
        \shortstack[r]{Example \\ architectures} & 
        \begin{minipage}{.1\textwidth}
        \includegraphics[height=32mm, trim=10 0 300 0, clip]{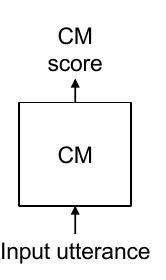}
        \end{minipage}
        &
        \begin{minipage}{.4\textwidth}
        \includegraphics[height=32mm, trim=110 0 0 0, clip]{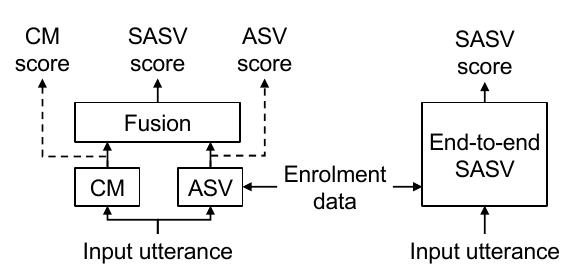}
        \end{minipage}
        \\
        Submitted scores & CM scores & SASV scores, optional CM \& ASV sub-system scores \\
        \bottomrule
    \end{tabular}
    }
\end{table*}

\section{Challenge outline}
\label{sec:outline}

We provide a brief description of the ASVspoof~5 database~\cite{wang_asvspoof5database}~(\S~\ref{sec:outline:database}), 
the stand-alone spoofed speech detection~(\S~\ref{sec:outline:t1}) and spoofing-robust ASV~(\S~\ref{sec:outline:t2}) challenge tracks, and both closed and open evaluation conditions~(\S~\ref{sec:outline:conditions}).\footnote{\textcolor{coloredit}{Following the previous edition of the ASVspoof challenge~\cite{liuASVspoof2021Spoofed2023}, we no longer separate the track for system ensembles from that for single systems.}}\footnote{Additional rules and participant guidelines not covered here are available in the challenge evaluation plan~\cite{ASVspoof5_evalplan_phase2}.}
Last, we describe the challenge baselines for the closed condition~(\S~\ref{sec:outline:baseline}) of each Track.

\begin{table}[!t]
    \centering
    \setlength{\tabcolsep}{4pt}
    \caption{Key ASVspoof~5 database statistics. Numbers in brace refer to target speakers relevant to Track~2 only.}
    \vspace{-3mm}
    \begin{tabular}{rllllc}
        \toprule
        
                  & \multicolumn{2}{c}{\#. speaker} & \multicolumn{2}{c}{\#. utterances} & \multirow{2}{*}{\shortstack{\#. attack}}\\
        \cmidrule(lr){2-3}\cmidrule(lr){4-5}
         & Female & Male & Bona fide & Spoofed &  \\  
        \midrule
        Train & 196 & 204 & \phantom{0}18,797 & 163,560 & \phantom{0}8\\
        Development & 392 (196) & 393 (202) & \phantom{0}31,334 & 109,616 & \phantom{0}8\\
        Eva. Track 1 & 370 & 367  &  138,688 & 542,086 & 16 \\
        Eva. Track 2 & 370 (194) & 367 (173) & 100,708 & 395,924 & 16\\
        \bottomrule
    \end{tabular}
    \label{tab:database_statistics}
\end{table}

\subsection{ASVspoof~5 Database}
\label{sec:outline:database}
ASVspoof~5 database~\cite{wang_asvspoof5database} statistics are presented in Table~\ref{tab:database_statistics}. 
Whereas previous ASVspoof databases are all generated using data collected from $\sim$100 speakers in highly controlled, studio-quality recording conditions, the ASVspoof~5 database is constructed from the English partition of the Multilingual Librispeech (MLS) database~\cite{pratap20_interspeech} which contains crowdsourced data collected from almost 2,000 speakers, each using their own acoustic and recording setup. 
Its crowdsourced nature ensures far greater variability than all previous ASVspoof databases.
Training, development, and evaluation sets are speaker-disjoint.
The training and development sets provide approximately 20k and 32k bona fide utterances, while there are in the order of 140k and 100k bona fide utterances in the evaluation sets.

The ASVspoof~5 database contains spoofing attacks generated using TTS and VC techniques, as well as adversarial attacks~\cite{panariello23b_interspeech,malacopula} for the first time. 
The set of TTS and VC attacks include contemporary algorithms (e.g., diffusion models~\cite{popov2021grad,popov2021diffusion}) as well as a legacy unit-selection system~\cite{steiner2018creating}. 
Attacks in the training, development, and evaluation sets are disjoint.  
Among the 16 attacks in the evaluation set, seven are adversarial attacks designed to manipulate the CM, ASV system, or both. 
They are referred to by attack identifiers from A01 to A32, with full details of each being provided in~\cite{wang_asvspoof5database}.
There are approximately 163k and 109k spoofed utterances for the training and development sets and in the order of 542k and 395k for the evaluation sets.

To study the impacts upon detection performance, a portion of bona fide and spoofed utterances in only the evaluation set are encoded or compressed using MP3, opus, amr, speex, m4a, a DNN-based tool called Encodec~\cite{de2023encodec}, the combination of MP3 and Encodec, or the simulated effects of transmission from a mobile device across a public switched telephone network. Full details are available in~\cite{wang_asvspoof5database}.

\subsection{Track 1}
\label{sec:outline:t1}
As illustrated by example architectures in Table~\ref{tab:track-summary}, 
Track 1 involves a stand-alone spoof/deepfake speech detection task (\texttt{bonafide} versus \texttt{spoof}).
It supports the evaluation of detection in isolation from ASV, a task which dates back to the first ASVspoof challenge edition held in 2015~\cite{Wu-ASVspoof2015}. 
The goal is to study the generalization and robustness of spoof/deepfake detection for a broad range of applications, e.g., call centres, telephone fraud, forensics, social media disinformation, \emph{etc}, in many of which there is no ASV system.

Participants are tasked with the design of a CM which should assign a single real-valued detection score to a given utterance. 
Higher CM scores are associated with a higher chance that the input utterance is \emph{bona fide}. 
Evaluation metrics are listed to the left of Table~\ref{tab:track-summary} and are described in~\S~\ref{sec:metrics}.

\subsection{Track 2}
\label{sec:outline:t2}
Track 2 extends the focus of ASVspoof to scenarios in which ASV systems are protected against spoof/deepfake attacks.
Solutions, referred to as spoofing-robust ASV (SASV) systems, are able to compare an unlabelled input utterance to an enrolment utterance(s) in the voice of the claimed speaker identity (target). 
Unlike standalone CM systems, SASV systems are evaluated using a mix of \emph{three} trial types --- \texttt{targets} (bona fide utterances from target speakers), \texttt{non-targets} (bona fide utterances from non-target speakers), and \texttt{spoof} (spoofed utterances). 
SASV systems should accept \texttt{target} trials only. 

Track 2 participants can develop SASV systems of any custom/preferred architecture (tandem, score fusion, embedding fusion, end-to-end, \emph{etc}). 
The more typical score fusion and end-to-end architectures are illustrated to the right of Table~\ref{tab:track-summary}.
Using a reference ASV sub-system provided by the challenge organisers, participants may nonetheless focus upon the development of a CM only.
No matter the architecture, a single SASV score must be provided.
Where distinct CM and ASV systems are used, e.g., as for score fusion systems, separate scores can also be provided for additional analyses. 
Track 2 metrics listed to the middle right of Table~\ref{tab:track-summary} are described in~\S~\ref{sec:metrics}.

The evaluation set for Track 2 is a subset of the ASVspoof 5 evaluation set, excluding data compressed with non-telephony codecs --- the DNN-based Encodec encoder, MP3, M4a, and the combination of Encodec and MP3. 

\subsection{Closed and open conditions}
\label{sec:outline:conditions}
For all previous ASVspoof challenges, participants were required to use only data specified in challenge protocols and contained in the training and development partitions for system optimisation. 
However, in recent years, and in similar fashion to trends in other fields of speech research, the use of speech foundation models pre-trained using self-supervised learning~\cite{mohamedSelfSupervised2022} and massive quantities of (bona fide) speech data has been explored in the spoof/deepfake speech detection community. 
Their use
has been found to improve detection performance across a range of datasets~\cite{ssl_frontends_asvspoof2021, wav2vec_antispoofing2022, zhangAudio2024}. 

Despite their appeal, the use of foundation models can undermine evaluation integrity since they can be trained using the same data used in generating spoofed data.  
Nonetheless, with the use of foundation models becoming the norm, the avoidance of data overlap in challenge and protocol design is becoming increasingly difficult.
In reality, it is practicably feasible, or even likely, that both attacks and defences will be optimised using common data resources. 
Since speech foundation models leverage massive quantities of data to train strong, often generic speech models having an enormous number of parameters, 
it is hardly a surprise that their use typically results in better performance than models trained using smaller data sets. 
Performance comparisons made between systems designed with or without the use of foundation models, as well as comparisons made between systems designed with the use of different foundation models are hence unfair. 
Accordingly, to protect evaluation integrity, while also supporting the use of foundation models, closed and open evaluation conditions were defined for both ASVspoof~5 tracks. 

The {\bfseries closed condition} follows the conventions of previous ASVspoof challenges and mandates use of only the ASVspoof~5 training partition for system training and the development partition for validation. For track 2, use of the Voxceleb2~\cite{voxceleb2} dataset was permitted for the training of SASV systems, or distinct ASV sub-systems. 

For the {\bfseries open condition} use of models pre-trained using external data was permitted, so long as there is no overlap with data contained in, or used in the generation of utterances contained in the ASVspoof~5 evaluation partition in terms of either speakers or utterances.\footnote{Compliant examples include SSL models trained using the LibriSpeech~\cite{panayotov2015librispeech} and VCTK~\cite{yamagishiCSTRVCTKCorpus2019} databases. Those pre-trained using LibriLight~\cite{kahnLibriLight2020}, however, are non-compliant since this database contains data collected from speakers included in the ASVspoof~5 evaluation partition. Further details are available in the ASVspoof~5 evaluation plan~\cite{ASVspoof5_evalplan_phase2}.} 
    The use of external data \emph{and} data within the ASVspoof~5 training partition was also permitted under the open condition.

\subsection{Baselines}
\label{sec:outline:baseline}
Baseline systems were defined for both tracks.
CM baselines for Track 1 include RawNet2~\cite{Jung2020,tak2021end} (B01) and AASIST~\cite{jung2022aasist} (B02).
Both CMs deliver competitive performance for previous ASVspoof challenge databases.
The pair of baselines for Track 2 are adopted from the SASV challenge~\cite{sasv2022}, and include an ASV-CM fusion-based system (B03) and an end-to-end system (B04). 
B03 uses a non-linear fusion~\cite{wangRevisiting2024} of the AASIST CM baseline B02 and an ECAPA-TDNN ASV system pre-trained using the VoxCeleb~2~\cite{voxceleb2} development partition. 
B04 is an end-to-end model~\cite{zhang2022mfa} which extracts embeddings from input and enrolment utterances and produces a single SASV score.\footnote{Implementations of all baseline systems 
are accessible from the ASVspoof~5 repository: https://github.com/asvspoof-challenge/asvspoof5}

\section{Metrics}
\label{sec:metrics}
In this section we summarize the performance metrics used for each of the two challenge tracks, as listed in Table~\ref{tab:track-summary}. 

\subsection{Track 1: from EER to DCF}
Following the familiar format of past challenge editions, Track 1 submissions were required to assign a real-valued detection score to each utterance.  Performance metrics were nonetheless revised to better reflect real-world operational CM applications. The relevant considerations are:
    \begin{itemize}
        \item detection threshold(s) must be set in advance;
        \item the miss and false alarm rates are not equally important. 
    \end{itemize}
The primary metric used previously for the assessment of standalone CMs --- the equal error rate (EER) --- is aligned with neither consideration. While use of the EER may be justified in pilot studies of bona fide-spoofed discrimination capability, its longer-term adoption risks overlooking design considerations relevant to the deployment of CMs in real-world applications.

Accordingly, the \emph{detection cost function} (DCF)~\cite{NIST-SRE-CTSplan-2020} metric was adopted for performance evaluation. While further details are available in~\cite{wang2024asvspoof5}, the DCF has the form
    \begin{equation}\label{eq:dcf-track1-norm}
        \text{DCF}(\tau_\text{cm}) = \beta \cdot P_\text{miss}^\text{cm}(\tau_\text{cm}) + P_\text{fa}^\text{cm}(\tau_\text{cm}),
    \end{equation}
where $P_\text{miss}^\text{cm}$ 
is the miss rate (false rejection rate of bona fide data) and $P_\text{fa}^\text{cm}$ 
is the false alarm rate (false acceptance rate of spoofed data). Both are functions of a detection threshold $\tau_\text{cm}$. The constant $\beta$ in \eqref{eq:dcf-track1-norm} is defined by $\beta:={C_\text{miss}(1-\pi_{\text{spf}})}/({C_\text{fa}\pi_{\text{spf}}})$ and is computed from pre-set costs for misses ($C_\text{miss}$) and false alarms ($C_\text{fa}$), as well as the spoofed and bona fide class priors ($\pi_\text{spf}$ and $1-\pi_\text{spf}$). 
Parameters for ASVspoof~5 give $\beta\approx 1.90$, i.e.\ missed detections of bona fide utterances are penalized nearly twice as much as false accepts of spoofed utterances~\cite{wang2024asvspoof5}.

The DCF in \eqref{eq:dcf-track1-norm} is used to compute both the \emph{minimum} and \emph{actual} 
DCF. The former, denoted minDCF, and the primary metric for Track~1, is the value of the DCF at the threshold that minimizes \eqref{eq:dcf-track1-norm} for evaluation data. 
The 
latter, denoted actDCF, uses a pre-set 
threshold $\tau_\text{Bayes}=-\log(\beta)$. 
Whereas minDCF measures performance using an `oracle' threshold (set according to ground-truth labels for evaluation data),
the actDCF is a measure of realised cost when the threshold is set \emph{before} observation of either evaluation data or labels. 

The reporting of both minDCF and actDCF provides complementary views of class discrimination (bona fide-spoof) and calibration (threshold setting generalization). A high actDCF could be due to either a lack of discrimination, calibration, or both --- it cannot be determined from the actDCF alone. The distinction between discrimination and calibration is important; whereas experimentation with alternative architectures to improve discrimination can be tedious and computationally demanding, calibration problems can, in principle, be addressed using relatively straightforward score-domain post-processing operations~\cite{ferrer_calibration_tutorial}. 
\textcolor{coloredit}{By definition, the actDCF is always greater than or equal to the minDCF of the corresponding system; the gap between these two values is affected by the goodness of calibration (see \cite[\S~2.5.2]{brummerBOSARIS2013} and discussion in the supplementary material).}

The $\tau_\text{Bayes}$ for actDCF is meaningful only when scores can be interpreted as calibrated log-likelihood ratios (LLRs)~\cite{Brummer2006-application-independent,ferrer_calibration_tutorial}.  
Similar to past challenge editions, the submission of LLR scores was not \emph{required} --- rather, it was \emph{encouraged} for the first time.\footnote{Readers unfamiliar with LLRs may rightfully wonder whether this requires modification of the model architecture. Following successful examples from speaker verification studies, this problem is typically addressed using a trainable calibration module (such as an affine transform) to post-process arbitrary detection scores into LLRs. Implementations such as~\cite{ferrer_calibration_tutorial} provide practical calibration recipes. Note, however, that any order-preserving score calibration does not affect the primary minDCF metric.} 
One important motivation to encourage the output of calibrated LLRs comes from the field of forensic voice comparison where evidence reporting through LLRs is well-established (e.g.~\cite{vanLierop2024-LLR-forensics}). 

In fact, one can measure the quality of arbitrary scores, in terms of their interpretation as calibrated LLRs. This can be accomplished using the \emph{cost of log-likelihood ratios} ($C_\text{llr}$)~\cite{Brummer2006-application-independent} metric used widely in speaker verification studies. 
The lower the $C_\text{llr}$, the better calibrated (and more discriminative) the scores. 
In addition to minDCF, actDCF, and $C_\text{llr}$, the EER is also reported so as to provide some consistency with previous challenge editions.

\subsection{Track 2: from SASV-EER to a-DCF}
For Track 2, participants could submit either single real-valued SASV scores 
or a triplet of scores 
which, in addition to SASV scores, contains 
spoof (CM sub-system) and speaker (ASV sub-system) detection scores. 
The former corresponds to any model architecture which outputs a single detection score, like for the end-to-end architecture illustrated to the lower right in Table~\ref{tab:track-summary}.
The latter assumes 
some appropriate fusion of CM and ASV 
scores~\cite{Kinnunen2020-tDCF} following the fusion architecture illustrated in Table~\ref{tab:track-summary}.

For both types of architecture, SASV 
scores are used to compute the 
\emph{normalized architecture-agnostic} detection cost function (a-DCF)~\cite{shim2024adcf}:
\begin{equation}
\begin{aligned}\label{eq:aDCF}
\text{a-}\text{DCF}(\tau_\text{sasv}) = &\alpha P_\text{miss}^\text{sasv}(\tau_\text{sasv}) + (1-\gamma) P_\text{fa,non}^\text{sasv}(\tau_\text{sasv}) \\
&+ \gamma P_\text{fa,spf}^\text{sasv}(\tau_\text{sasv}),
\end{aligned}
\end{equation}
where $P_\text{miss}^\text{sasv}$ 
is the ASV miss (false rejection of target speakers) rate and where $P_\text{fa,non}^\text{sasv}$ 
and $P_\text{fa,spf}^\text{sasv}$ 
are the false acceptance rates for non-target and spoof attack trials respectively. 
All three error rates are functions of a single SASV threshold $\tau_\text{sasv}$, and the constants $\alpha$ and $\gamma$ are 
obtained from detection costs and priors, with 
values $\alpha \approx 1.58$ and $\gamma \approx 0.84$~\cite{wang2024asvspoof5}.
The primary metric for Track 2 is the minimum a-DCF, obtained as the a-DCF at the threshold that minimizes~\eqref{eq:aDCF} for evaluation data.
    
The \emph{ASV-constrained minimum tandem detection cost function} (t-DCF)~\cite{Kinnunen2020-tDCF} and the \emph{tandem equal error rate}~(t-EER) \cite{Kinnunen2024-tEER} are also reported for submissions which provide distinct ASV and CM sub-system scores.
The ASV-constrained t-DCF, the primary metric since ASVspoof 2019, is computed using the same costs and priors as the a-DCF and using ASV scores produced by a common ASV system (that of B03) in place of scores provided by the participant. 

The t-EER can be seen as a generalisation of the conventional two-class, single system EER which provides an application-agnostic discrimination measure.
For computation of the t-EER, both CM and ASV sub-system scores are used to obtain a single \emph{concurrent t-EER} value. It has a simple interpretation as the error rate for the unique \emph{pair} of ASV and CM thresholds at which the miss rate and the two types of false alarm rate (one non-target, the other for spoofing attack trials) are equal~\cite{Kinnunen2024-tEER}.

\begin{figure*}[t!]
    \centering
    \subfloat[Track 1 closed condition]{\includegraphics[width=\textwidth, trim=0 10 0 0]{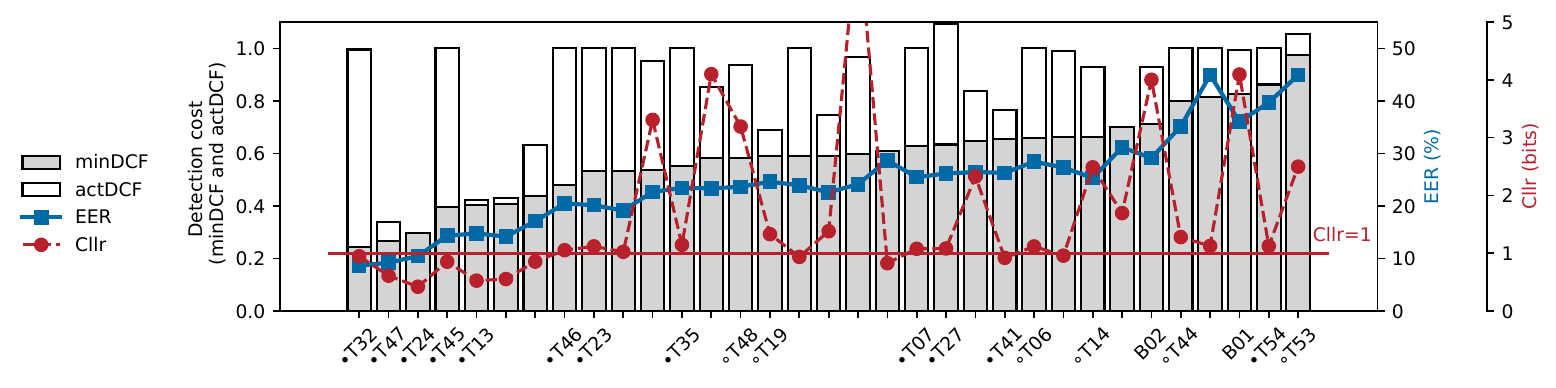}\label{fig:result_all_t1_closed}}
    \vspace{-3mm}
    \hfill
    \subfloat[Track 1 open condition]{\includegraphics[width=\textwidth, trim=0 10 0 0]{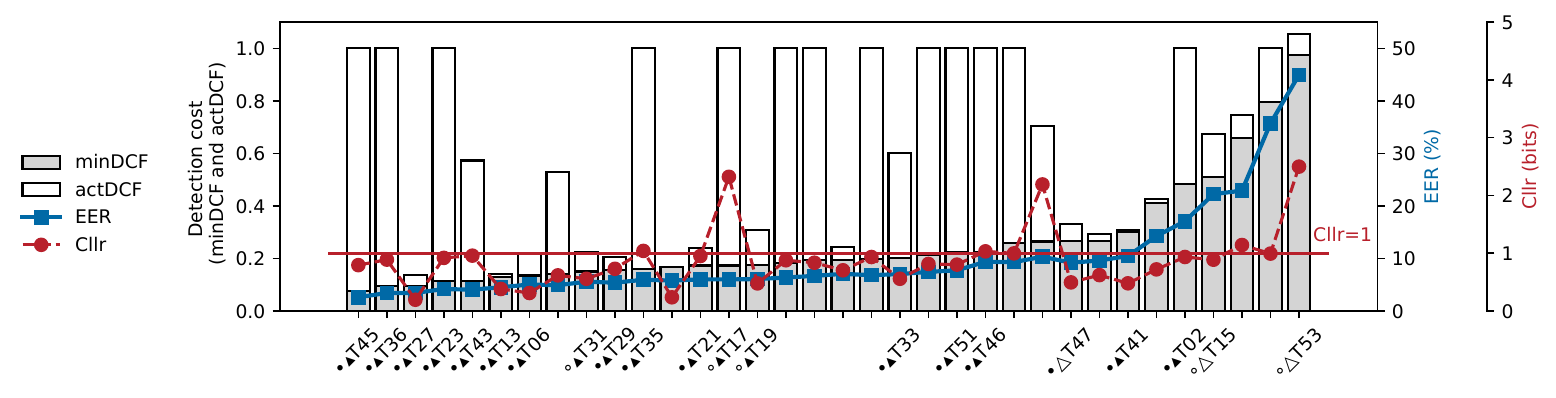}\label{fig:result_all_t1_open}}
    \caption{Results of ASVspoof~5 challenge Track 1. Ensemble and single systems are marked by $\bullet$ and $\circ$, respectively. Open-condition submissions using and not using pre-trained foundation models are marked by $\blacktriangle$ and $\triangle$, respectively. Note that a system's actDCF value is no smaller than its minDCF value. 
    }
    \label{fig:results_all_t1}
\end{figure*}

\section{Results and submissions}
\label{sec:results}

\newcommand{\augrb}{RB}
\newcommand{\augrr}{RV}
\newcommand{\augsp}{SP}
\newcommand{\augpp}{PP}
\newcommand{\featspc}{spec.}
\newcommand{\fusionwa}{W.avg}
\newcommand{\fusionlr}{Log.Reg.}

\begin{table*}[t!]
  \setlength{\tabcolsep}{2pt}
  \centering
  \caption{Summary of top submissions for each track. Submissions are presented in order according to results of the primary evaluation metric of each Track. 
  The symbol $\blacktriangle$ marks acoustic frontend using a pre-trained speech foundation model. 
  Abbreviations are defined for room reverberation (\augrr), RawBoost (\augrb), speed perturbation (\augsp), pitch perturbation (\augpp), spectrogram (\featspc), weighted average (\fusionwa), and logistic regression (\fusionlr). \textcolor{coloredit}{Details are shown for only three of the top submissions to the Track 2 closed condition for which system reports where received.}
  }
  \begin{tabular}{rrlllllc}
  \toprule
     &  & ID & Data Augmentation & Acoustic Frontend & Backend Classifier & Fusion (\#. sub-systems) & Ref.  \\ \midrule
    \multirow{15}{*}{\rotatebox{90}{Track 1}} & 
    \multirow{8}{*}{\rotatebox{90}{Closed}} & 
    T32 & Pre-emph., SpecAug, low-pass filter & Waveform & Transformer & Unknown (3)  & N/A \\  \cmidrule{3-8} 
    & & T47 & \begin{tabular}{@{}l@{}}Noise, codec, \augrb,  vocoder, \augsp. \end{tabular} & \begin{tabular}{@{}l@{}}Mel \featspc\end{tabular} & ResNet & \fusionwa (10)  & \cite{tran24_asvspoof_t47} \\ \cmidrule{3-8} 
    & & T24 & \begin{tabular}{@{}l@{}} Noise, codec, \augrr, \augpp, \augsp \end{tabular} & \begin{tabular}{@{}l@{}}Waveform, mel \featspc\end{tabular} & \begin{tabular}{@{}l@{}}ResNet, AASIST, ConvViT-Base\end{tabular} & \fusionlr (3) & \cite{duroselle24_asvspoof_t24} \\ \cmidrule{3-8}
    & & T45 & Vocoder, codec & Waveform & RawNet2, AASIST & \fusionwa (4) & \cite{chen24_asvspoof_t45} \\ \cmidrule{3-8}
    & & T13 & \begin{tabular}{@{}l@{}}Codec, \augrb, \augrr, \augsp \end{tabular} & Waveform & AASIST & Average (4) & N/A \\ 
    \cmidrule{2-8}
    & \multirow{8}{*}{\rotatebox{90}{Open}} & 
    T45 & Vocoder, codec, TTS, noise, \augrr & $\blacktriangle$wav2vec2-large & GAT, MFA-Res2Net, LSTM & \fusionwa (6) & \cite{chen24_asvspoof_t45} \\ \cmidrule{3-8}
    & & T36 & \augrb, noise, high/low-pass filtering & $\blacktriangle$WavLM-Base & MLP & Average (5) & \cite{aliyev24_asvspoof_t36} \\ \cmidrule{3-8}
    & & T27 & \begin{tabular}{@{}l@{}}Noise, codec (mp3, ogg) \augrr \end{tabular} & $\blacktriangle$WavLM-Base & MHFA, WAP & \fusionlr (3) & \cite{stourbe24_asvspoof_t27} \\ \cmidrule{3-8}
    & & T23 & \begin{tabular}{@{}l@{}}Silence trim., noise, SpecAug, \augrb \\\augsp, \augpp, \augrr, codec \end{tabular} & \begin{tabular}{@{}l@{}}LFCC, $\blacktriangle$wav2vec2-large\end{tabular} & \begin{tabular}{@{}l@{}}LCNN, GNN, Conformer\end{tabular} & Median pooling (3) & \cite{falez24_asvspoof_t23} \\ \cmidrule{3-8}
    & & T43 & \begin{tabular}{@{}l@{}}Time-mask, noise, \augrr, \augrb, codec\end{tabular} & \begin{tabular}{@{}l@{}}$\blacktriangle$wav2vec2-large\end{tabular} & \begin{tabular}{@{}l@{}} AASIST\end{tabular} & Average (2) & \cite{xu24_asvspoof_t43} \\
    \midrule
    \multirow{20}{*}{\rotatebox{90}{Track 2}} & 
    \multirow{6}{*}{\rotatebox{90}{Closed}} & 
    T45 & \begin{tabular}{@{}l@{}}Vocoder, codec, noise, \augrr, \augsp \end{tabular} & \begin{tabular}{@{}l@{}} CM: Waveform \\ ASV: mel \featspc \end{tabular} & \begin{tabular}{@{}l@{}}CM: RawNet2, AASIST \\ ASV: ResNet240 \end{tabular} & \begin{tabular}{@{}l@{}} \fusionwa{} of CMs (CM 12)  \\ Rule for ASV+CM (ASV 1) \end{tabular} & \cite{chen24_asvspoof_t45} \\ \cmidrule{3-8}
    & & T47 & \begin{tabular}{@{}l@{}}Noise, \augrb, codec, vocoder, \augsp \end{tabular} & \begin{tabular}{@{}l@{}}Mel \featspc\end{tabular} & \begin{tabular}{@{}l@{}} CM: ResNet \\ ASV: ResNet152, ResNet293 \end{tabular} & \begin{tabular}{@{}l@{}}\fusionwa{} of all \\ (ASV 2, CM 10) \end{tabular} & \cite{tran24_asvspoof_t47} \\ \cmidrule{3-8}
    & & T24 & \begin{tabular}{@{}l@{}}Noise, \augrr, codec, \augpp, \augsp \end{tabular} & \begin{tabular}{@{}l@{}}CM: Waveform, mel \featspc \\ ASV: mel \featspc \end{tabular} & \begin{tabular}{@{}l@{}} CM: ResNet, AASIST, ConvViT-Base \\ ASV: ResNet34 \end{tabular} & \begin{tabular}{@{}l@{}}  \fusionlr{} for CMs (CM 3) \\ LLR-fusing ASV\&CM (ASV 1)\end{tabular} & \cite{duroselle24_asvspoof_t24} \\ 
    \cmidrule{2-8}
    & \multirow{12}{*}{\rotatebox{90}{Open}} & 
    T45 & \begin{tabular}{@{}l@{}}Noise, \augrb, \augsp,  codec\end{tabular} & \begin{tabular}{@{}l@{}} CM: $\blacktriangle$wav2vec2-large \\ ASV: mel \featspc \end{tabular} & \begin{tabular}{@{}l@{}} CM: GAT, MFA-Res2Net, LSTM \\ ASV: ResNet240 \end{tabular} & 
    \begin{tabular}{@{}l@{}} Same as T45 in closed cond. \\ (ASV 1, CM 12) \end{tabular} & \cite{chen24_asvspoof_t45} \\ \cmidrule{3-8}
    & & T39 & \begin{tabular}{@{}l@{}}SpecAug, \augrr, noise \end{tabular} & \begin{tabular}{@{}l@{}}CM: $\blacktriangle$wav2vec2, \\ $\quad\quad$ Data2Vec \\ ASV: mel \featspc \end{tabular} & \begin{tabular}{@{}l@{}}CM: ResNet100, ReDimNet-B2 \\ ASV: ResNet100 \end{tabular} & \begin{tabular}{@{}l@{}} \fusionwa{} for CMs (CM 6) \\ min of ASV \& CM score (ASV 1) \end{tabular} & \cite{okhotnikov24_asvspoof_t39} \\ \cmidrule{3-8}
    & & T36 & \begin{tabular}{@{}l@{}}\augrb, \augrr, noise, \augsp \end{tabular} & \begin{tabular}{@{}l@{}} CM: $\blacktriangle$WavLM-base \\ ASV: mel \featspc \end{tabular} & \begin{tabular}{@{}l@{}} CM: MLP \\ ASV: ResNet \end{tabular}& \begin{tabular}{@{}l@{}} \fusionwa{} for CMs (CM 5) \\ $\text{CM}^{1000} * \text{ASV}$ (ASV 1) \end{tabular} & \cite{aliyev24_asvspoof_t36} \\  \cmidrule{3-8}
    & & \textcolor{coloredit}{T06} & \begin{tabular}{@{}l@{}}Silence trim., vocoder, \augrb \end{tabular} & \begin{tabular}{@{}l@{}} CM: $\blacktriangle$wav2vec2-large, \\ $\quad\quad$ WavLM-base \\ ASV: Waveform \featspc \end{tabular} & \begin{tabular}{@{}l@{}} CM: MLP \\ ASV: TitaNet \end{tabular}& \begin{tabular}{@{}l@{}} Average for CMs (CM 2) \\ LLR-fusion ASV\&CM (ASV 1) \end{tabular} & \cite{martindonas24_asvspoof} \\ 
    \cmidrule{3-8}
    & & \textcolor{coloredit}{T29} & \begin{tabular}{@{}l@{}}\augrr, noise,  \end{tabular} & \begin{tabular}{@{}l@{}} CM: $\blacktriangle$WavLM-base \\ ASV: Waveform \featspc \end{tabular} & \begin{tabular}{@{}l@{}} CM: MLP \\ ASV: ECAPA-TDNN \end{tabular}& \begin{tabular}{@{}l@{}} N/A (CM 1) \\ LLR-fusion ASV\&CM (ASV 1) \end{tabular} & N/A \\ 
    \bottomrule
  \end{tabular}
\label{tab:system_summary}
\end{table*}

\begin{figure}
    \centering
    \subfloat[Attacks in closed condition]
    {\includegraphics[width=0.5\textwidth, trim=0 5 0 17, clip]    {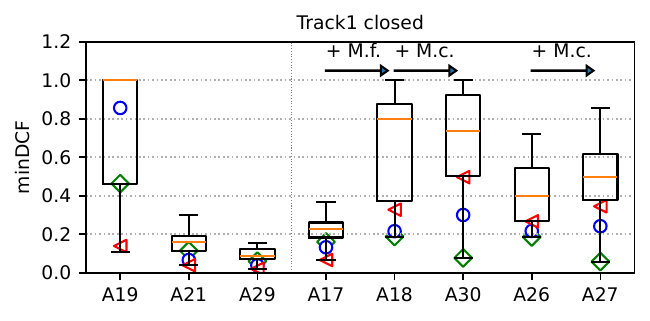}
    \label{fig:ana_selected_attacks_t1_closed}
    }
    \hfill
    \subfloat[Attack groups in closed (left) and open (right) conditions]
    {\includegraphics[width=0.5\textwidth, trim=0 5 0 7, clip]    {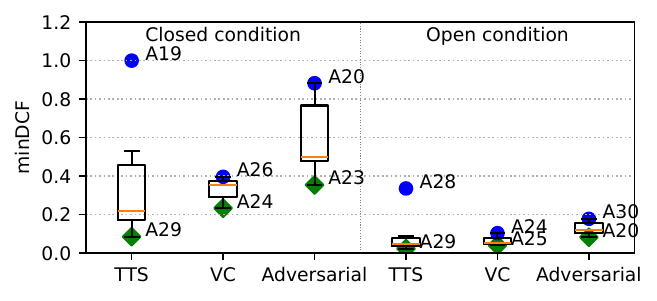}
    \label{fig:ana_merged_attacks_t1}
    }
    \hfill
    \subfloat[Codec groups in closed (left) and open (right) conditions]
    {\includegraphics[width=0.5\textwidth, trim=0 5 0 7, clip]    {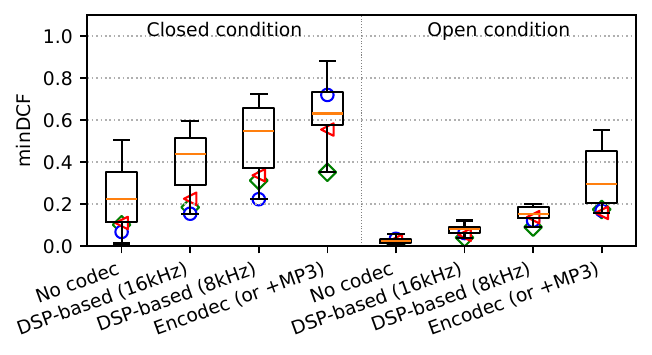}
    \label{fig:ana_merged_codec_t1}
    }
    \caption{Boxplots of evaluation set minDCF of Track 1. 
    In sub-figure (a), each box shows the raw minDCF values of top 50\% submissions in the closed condition. Markers are top-1 submission (\textcolor{teal}{$\diamond$}), top-2 (\textcolor{blue}{o}), and top-3 (\textcolor{red}{$\triangleleft$}) submissions. 
    The annotated arrows `+ M.f.' and `+ M.c.' mean that attacks are the right hand side are obtained via applying Malafide annd Malacopula, respectively, to the attacks on the left hand side. 
    Figures for other tracks and conditions are presented in the appendix.
    In sub-figure (b), the median minDCF value of the top 50\% submissions for each attack is computed, and each box summarizes the median minDCF values of the attacks in the group (either TTS, VC, or adversarial). Markers are easiest (\textcolor{teal}{$\blacklozenge$}) and hardest (\textcolor{blue}{$\bullet$}) attacks. 
    In sub-figure (c), each box shows the raw minDCF values of top 50\% submissions in a codec condition. Markers are the same as (a). \textcolor{coloredit}{Orange lines denote the median  minDCFs.}
    }
    \label{fig:ana_t1}
\end{figure}

In the following we present results for each track and each condition.  Also provided is a summary of top-ranked submissions and principal findings.

\subsection{Track 1}

\subsubsection{Closed condition} Results are illustrated in Figure~\ref{fig:results_all_t1}\subref{fig:result_all_t1_closed}. Submissions\footnote{Submissions without a team identifier correspond to teams that did not submit a valid system description. As per ASVspoof Challenge policy, neither the team name nor the names of team members can be revealed.} are ranked according to performance for evaluation data and the primary minDCF metric (gray bars).
Most submissions outperform the baselines, with 27 teams beating the best B02 baseline.
Whereas the \textit{T32} submission achieves the lowest minDCF and EER (blue squares), the lowest $C_\text{llr}$ (red circles) is obtained for the \textit{T24} submission, 
indicating better \emph{goodness}~\cite{nautsch2019speaker} of the scores for making Bayes decisions given different priors and decision costs. 
The lowest $C_\text{llr}$ for the \textit{T24} submission corresponds to the lowest actDCF, an indication of strong detection performance at the Bayes threshold for organizer-specified priors and decision costs. 
\textcolor{coloredit}{The gap between the actDCF and minDCF for \textit{T24} is negligible, supporting the finding that the pre-defined threshold yields near-optimal decision performance given the scores produced by \textit{T24}.}

The variation in EER and $C_\text{llr}$ \textcolor{coloredit}{as well as the gap between actDCF and minDCF}, all shown in Figure~\ref{fig:results_all_t1}\subref{fig:result_all_t1_closed}, demonstrate that systems with strong discrimination performance (i.e., with low EER and minDCF) cannot necessarily make useful Bayes decisions. 
Systems for which the $C_\text{llr}$ is equal to or higher than 1 bit perform no better than a random coin toss\textcolor{coloredit}{, and the decisions `are better made by omitting these systems'~\cite[\S2.4.7]{nautsch2019speaker}.  
Ideally, systems should yield a $C_\text{llr}$ of less than 1 bit to be useful for Bayes decision-making.}

A summary of top-performing systems is presented to the top of Table~\ref{tab:system_summary}. To facilitate comparisons, systems are decomposed into four major components that define the training and inference pipeline: data augmentation, the acoustic frontend, backend classifier, and sub-system fusion. 
In terms of data augmentation, the best-performing systems for the closed condition rely primarily on digital signal processing (DSP) techniques (e.g., SpecAugment~\cite{park19e_interspeech}). A number of submissions also incorporated RawBoost~\cite{rawboost}, codec compression, and speed perturbation. 
Perhaps unsurprisingly, there is no use of SSL frontends, quite possibly due to the lack of sufficient training data permitted under the closed condition.
Instead, the dominant acoustic representation is mel spectrograms processed typically using deep neural classifiers such as ResNet~\cite{he2016deep}, raw waveform inputs like for AASIST~\cite{jung2022aasist}, or hybrid architectures combining convolutional networks and vision-transformer modules (e.g., ConvViT-Base). Finally, most submissions are ensemble systems, with fusion strategies typically combining three-to-four subsystems using logistic regression or score-level averaging.

A summary of results for a selection of 8 specific spoofing attacks\footnote{With full descriptions being available in~\cite{wang_asvspoof5database},
we provide here only essential details of specific attack algorithms.  Results for the full complement of attack algorithms are available in the appendix.} is shown in Figure~\ref{fig:ana_t1}\subref{fig:ana_selected_attacks_t1_closed}.
Boxplots illustrate the distribution in minDCF for the top 50\% of submissions, while results for the top 3 systems are illustrated by coloured markers.
The most challenging attack is that of A19, the concatenative MaryTTS system~\cite{schroder11interspeech}.
The lowest minDCFs are obtained for attacks A21 and A29, both contemporary zero-shot TTS systems~\cite{wang_asvspoof5database}. \textcolor{coloredit}{In the case of A19, CM systems trained using DNN-based spoof/deepfake data may fail to capture waveform concatenation artifacts. For example, abrupt changes in the fundamental frequency (F0) caused by waveform concatenation~\cite{kirchhubel22_interspeech} are unlikely to be present in DNN-based spoof/deepfakes (e.g., A01 in the training set) which generate speech with a smooth F0 trajectory.} Thus, robust performance for relatively advanced attacks is no guarantee of protection against attacks implemented with legacy technology.

The 5 right-most box plots in Figure~\ref{fig:ana_t1}\subref{fig:ana_selected_attacks_t1_closed} illustrate the impact of adversarial attacks applied to the base A17 zero-shot TTS system and the base A26 zero-shot VC system.
For the former, the Malafide attack~\cite{panariello23b_interspeech} provokes a substantial increase in the minDCF for attacks A18.
The Malacopula attack~\cite{malacopula}, when applied either alone to attack A26 (giving A27) or in combination with Malafide to attack A17 (giving A30), is also damaging, albeit to a lesser extent. 
This is not entirely surprising given that, while Malafide targets the manipulation of spoof/deepfake detection systems, Malacopula targets ASV systems, whereas Track~1 concerns spoof/deepfake detection only. 
\textcolor{coloredit}{Malafide and Malacopula are implemented using digital filters whose coefficients are updated via gradient descent. The optimisation objective is to increase the false acceptance rates of proxy CM and ASV systems, respectively. Consequently, these results suggest that the adversarial post-processing of spoof/deepfake attacks can increase the decision error rates of many CM systems.}

\subsubsection{Open condition}  Results are illustrated in Figure~\ref{fig:results_all_t1}\subref{fig:result_all_t1_open}.
As expected, minDCF and EER values are lower than for the closed condition, reflecting the benefit of large, pretrained SSL models. 
Despite lower minDCF results, some of the top systems obtain higher actDCF values close to 1.0 and $C_\text{llr}$ values close to 1 bit, suggesting poor calibration.
In contrast, the $C_\text{llr}$ of 0.2 for the \textit{T27} system indicates both strong discrimination and calibration performance.

Table~\ref{tab:system_summary} shows no substantial differences in the use of data augmentation for the open condition.
Large foundational models in the form of SSL-based architectures such as wav2vec~2.0~\cite{Baevski2020} and WavLM~\cite{chen2022wavlm} acoustic frontends dominate and are  
fine-tuned jointly with a backend classifier.  The strong representational capacity of SSL frontends leads to the use of relatively lightweight backend architectures, e.g.\ multi-layer perceptrons (MLPs) and LCNNs.
System fusion involves two-to-six subsystems, with weighted score averaging being the most common strategy.

A picture of the improvements in detection performance for the open vs.\ closed conditions is presented in Figure~\ref{fig:ana_t1}\subref{fig:ana_merged_attacks_t1}.
Boxplots illustrate the distribution in minDCF for TTS, VC and adversarial attacks for the top 50\% of submissions.
The easiest and most difficult attacks are illustrated in each case.
Improvements to the minDCF for the open condition \textcolor{coloredit}{are} substantial for all three attack classes and the gap between them is greatly reduced, including for adversarial attacks, even if minDCFs remain generally higher than for TTS and VC attacks. 
Unlike for the closed condition, the legacy A19 attack is among the easiest to detect.\footnote{Results shown in the appendix.} The most challenging to detect is A28, a pre-trained zero-shot YourTTS~\cite{casanova2022yourtts} system released with the Coqui toolkit~\cite{Eren_Coqui_TTS_2021}, for which the minDCF is 0.33.

\subsubsection{Influence of codecs and compression}
A similar picture of comparative performance for open and closed conditions with respect to the encoding and compression schemes is presented in Figure~\ref{fig:ana_t1}\subref{fig:ana_merged_codec_t1} in the form of minDCF boxplots for the top 50\% of submissions. DNN-based Encodec compression and its combination with MP3 are the most challenging, followed by narrow band 8~kHz DSP-based codecs, then 16~kHz DSP-based codecs. The top-1 submission in the closed condition is substantially better (minDCF=0.35) than the second best submission in the case of Encodec (minDCF=0.55). 
The improvement in minDCF for open conditions is substantial. 
For Encodec, the top-3 submissions achieve a minDCF value below 0.2, and the median minDCF of the top 50\% submissions is 0.26. In other cases, the median minDCF is below 0.2.

\subsection{Track 2}
\label{sec:results:t2}
In the following we present a summary of Track 2 results. Visualizations of performance for individual attacks, attack types, and the influence of codecs and compression can be found in the appendix.

\begin{figure*}
    \centering
    \subfloat[Track 2 closed condition]{\includegraphics[width=0.5\textwidth, trim=0 10 0 0]{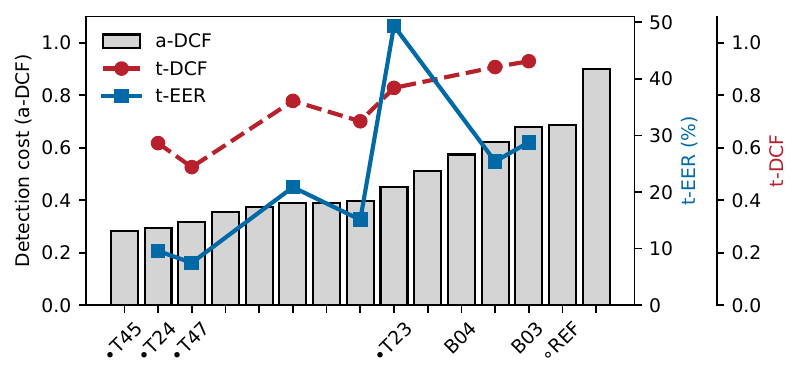}\label{fig:result_all_t2_closed}}
    \subfloat[Track 2 open condition]{\includegraphics[width=0.5\textwidth, trim=0 10 0 0]{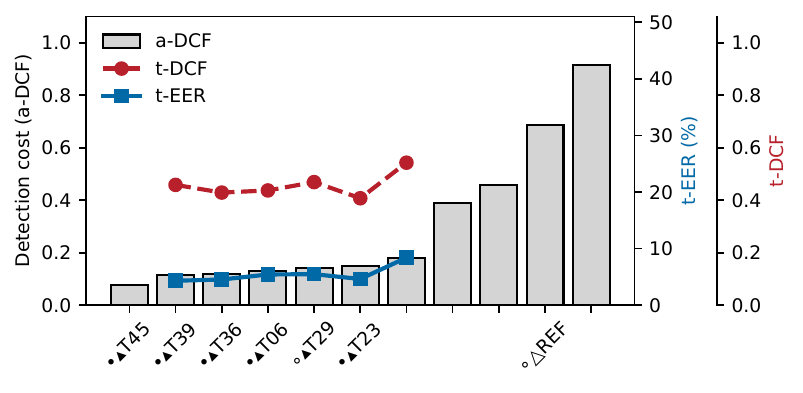}\label{fig:result_all_t2_open}}
    \caption{Results of ASVspoof~5 challenge Track 2. Ensemble systems and single systems are marked by $\bullet$ and $\circ$, respectively. Open-condition submissions using and not using pre-trained self-supervised models are marked by $\blacktriangle$ and $\triangle$, respectively. System REF refers to the organisers’ ASV without a CM. Results of t-DCF and t-EER are presented if the system submitted the optional CM and ASV scores. 
    }
    \label{fig:results_all_t2}
\end{figure*}

\subsubsection{Closed condition} Results for the closed condition are presented in Figure~\ref{fig:results_all_t2}\subref{fig:result_all_t2_closed}. 
Submissions are ranked according to the min a-DCF for evaluation data (gray bars).
More than half of submissions outperform the best baseline B04 as well as the organisers' ASV system without a CM sub-system~(REF). The \textit{T45} submission achieves the lowest min a-DCF of 0.28. 
Among submissions for which separate ASV and CM scores were both provided, the \textit{T47} submission achieves the lowest t-EER (blue squares) of 7.49\% and t-DCF (red circles) of 0.53, followed by \textit{T24}. 
Note that both the t-EER and t-DCF reflect detection performance for submissions having tandem ASV and CM sub-systems, while the min a-DCF reflects the detection performance of systems which provide only a single score (such as those produced from the fusion of separate ASV and CM scores).
Results hence show that the ranking of tandem ASV and CM systems, as in the case of submissions \textit{T47} and \textit{T24},
can differ when ranking is instead performed using fused scores.

A summary of top-performing systems is presented to the middle of Table~\ref{tab:system_summary}. 
The augmentation techniques are similar to those used for Track 1 closed condition submissions and include RawBoost, speed perturbation, and other DSP-based techniques. 
The top 3 systems use separate ASV and CM sub-systems, with the number of CM sub-systems being consistently larger than the number of ASV sub-systems. 
Participants focused their efforts upon robustness to spoofing rather than ASV, an indication that there is more to be gained from optimising the former than the latter. 
There is comparatively little variation in ASV system architectures, with mel-spectrograms being the preferred acoustic frontend, and ResNet-based models being the dominant backend classifier. 
There is substantial variation in fusion strategies, from simple linear averaging to non-linear methods such as~\cite{wangRevisiting2024}. 

A performance analysis\footnote{See 
Figure~6(a)
in the appendix.} 
for the same 8 spoofing attacks as in Figure~\ref{fig:ana_t1}\subref{fig:ana_selected_attacks_t1_closed} shows trends consistent with those for the Track 1 closed condition.
The only exception is Malacopula which, when applied to A26 (giving A27) or in combination with Malafide to attack A17 (giving A30), provokes an increase of more than 0.1 in the median min a-DCF for the top 50\% of submissions. 
This is expected since Malacopula targets the ASV system. 
As for the Track 1 closed condition, the concatenative MaryTTS attack A19 remains the most challenging to detect.

\subsubsection{Open condition} Results for the open condition are presented in Figure~\ref{fig:results_all_t2}\subref{fig:result_all_t2_open}. 
The use of SSL-based foundation models again leads to considerably better results. 
The \textit{T45} submission achieves a min a-DCF of 0.07, while the 2nd to the 5th ranked systems achieve min a-DCF values between 0.11 and 0.14. 

System summaries shown to the bottom of Table~\ref{tab:system_summary} show that most of the top teams reused the same CM architectures used for their corresponding submissions to the Track 1 open condition, for which the same teams also rank among the top performers.
\textcolor{coloredit}{For example, the \textit{T45} submission used the same CM architecture for both Track~1 and Track~2, combining a wav2vec2 Large acoustic frontend with a GAT, MFA-Res2Net, and LSTM backend classifier. The second-ranked \textit{T36} submission used WavLM-Base and MLP for both tracks.}
Again for the open condition, the number of CM sub-systems is substantial, varying from 3 to~12. 

A deeper analysis of results\footnote{See 
Figure~6(b)
in the appendix.} 
shows similar trends to those for Track 1 illustrated in Figure~\ref{fig:ana_t1}\subref{fig:ana_merged_attacks_t1}. 
Improvements to the min a-DCF for the open conditions are again substantial for the three types of attacks and the gap in performance for each type is greatly reduced. 
One notable difference is that the easiest and most difficult adversarial attacks to detect for the open condition become A18 and~A30. 
This difference is again expected because A18 is the product of an easily-detectable TTS attack (A17) and the Malafide attack which targets spoofing detection systems, whereas A30 is the combination of A18 and Malacopula attacks which target ASV systems.
Like for the Track 1 open condition, the most challenging attack to detect is A28.

\begin{figure*}
    \centering
    \subfloat[\textit{T45} of Track 1 open condition]{\includegraphics[width=0.45\textwidth, trim=0 5 0 0, clip]{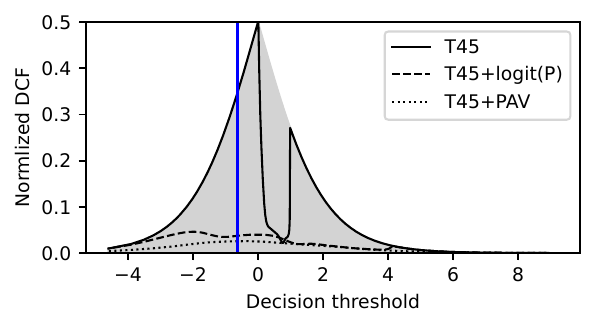}
    \label{fig:ana_ape_1}
    }
    \subfloat[\textit{T27} of Track 1 open condition]{\includegraphics[width=0.45\textwidth, trim=0 5 0 0, clip]{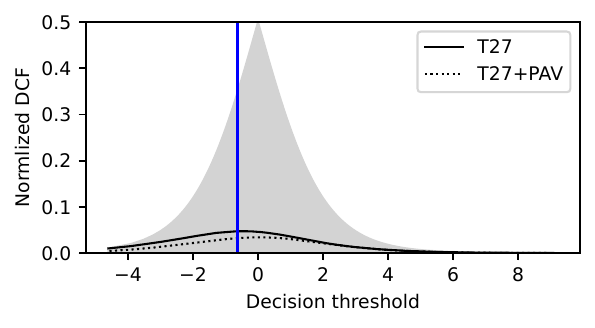}
    \label{fig:ana_ape_2}
    }
    \caption{Values of normalized DCF at different decision thresholds (\S~\ref{sec:dis:calibration_cm}). The blue vertical line marks the threshold for Track 1 actDCF computation. The shaded area is upper-bounded by the normalized DCF of a dummy CM that rejects or accept all trials.}
    \label{fig:ana_ape}
\end{figure*}
\begin{figure*}
\vspace{-5mm}
    \centering
    \subfloat[\textit{T45} of Track 1 open condition]{\includegraphics[width=0.45\textwidth, trim=0 5 0 15, clip]{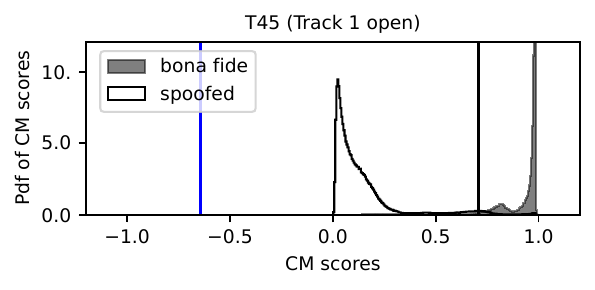}
    \label{fig:ana_scores_1}
    }
    \subfloat[\textit{T27} of Track 1 open condition]{\includegraphics[width=0.45\textwidth, trim=0 5 0 15, clip]{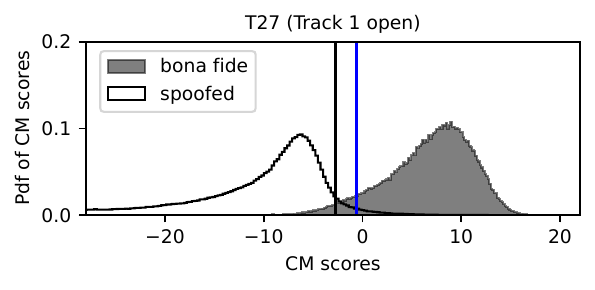}
    \label{fig:ana_scores_2}
    }
    \caption{Distributions of CM scores from submission \textit{T45} (left) and \textit{T24} (right) in Track 1 open condition. The blue and black vertical lines correspond to the Bayesian decision threshold and the one achieving the min DCF, respectively.}
    \label{fig:ana_score}
\end{figure*}

\section{Discussion}

\subsection{CM score calibration}
\label{sec:dis:calibration_cm}
Previous ASVspoof challenges have focused on evaluating the \emph{discrimination} power of submitted systems in terms of the EER or min t-DCF. 
Both metrics require the setting of an `ideal' decision threshold either so that the miss and false alarm rates are equal, or to minimise the t-DCF. 
In deployment, however, ground truth labels are obviously not available.
The decision threshold must instead be set by the system user, e.g., using asserted priors and application-dependent decision costs or by empirical optimisation using development data. 
User-supplied decision thresholds are unlikely to be `ideal'. 

Evaluating the \emph{calibration} power of a system gauges the goodness of its decision making capability across different applications (i.e., user-supplied decision thresholds). While the $C_\text{llr}$ (Section~\ref{sec:metrics}) summarizes system performance over `an infinite spectrum of operating points'~\cite{van2007introduction}, 
to illustrate the calibration issue more intuitively, we plot the decision errors of a system as a function of the decision threshold~\cite{brummer2021trials}.

We use the \textit{T45} and \textit{T27} submissions to the Track 1 open condition. The \textit{T45} system obtains the lowest minDCF (i.e., the best discriminative power) but performs much worse in terms of $C_\text{llr}$ and actDCF (i.e., supposedly due to poor calibration). In contrast, the \textit{T27} system performs well in both aspects. 
Given the scores produced by each system, we compute normalized DCF values\footnote{Following the error analysis in existing literature~\cite{brummer2021trials}, we use a normalized DCF, which is a scaled version of the DCF defined in (\ref{eq:dcf-track1-norm}): $\frac{C_{\text{fa}}\tilde{\pi}_\text{spf}}{C_{\text{fa}}\tilde{\pi}_\text{spf} + C_\text{miss}(1-\tilde{\pi}_\text{spf})}\text{DCF}(\tau_{\text{Bayes}}(\tilde{\pi}_\text{spf})) = \frac{1}{1+\beta(\tilde{\pi}_{\text{spf}})}\text{DCF}(\tau_{\text{Bayes}}(\tilde{\pi}_\text{spf}))$.} 
but use a spectrum of Bayes thresholds 
$\tau_{\text{Bayes}}(\tilde{\pi}_\text{{spf}}) =-\log\big(\beta(\tilde{\pi}_\text{{spf}})\big)$, where $\beta(\tilde{\pi}_\text{{spf}})=C_\text{miss}(1-\tilde{\pi}_\text{spf})/C_\text{fa}\tilde{\pi}_{\text{spf}}$ is computed using the challenge-specified decision costs ($C_\text{miss}$ and $C_\text{fa}$) and a spoofed class prior $\tilde{\pi}_{\text{spf}}$ varying from 0.001 to 0.999. 
The black solid curve in Fig.~\ref{fig:ana_ape}\subref{fig:ana_ape_1} illustrates the normalized DCF values for the \textit{T45} system as a function of $\tau_{\text{Bayes}}(\tilde{\pi}_{\text{spf}})$. 
For reference, the shaded area is upper bounded~\cite{brummer2021trials} by the decision cost of a dummy system which either rejects or accepts all the trials, whichever is lower.

Interestingly, \textit{T45} hits the upper bound across many decision thresholds, including that used for actDCF illustrated by the vertical blue line in Fig.~\ref{fig:ana_ape}\subref{fig:ana_ape_1}). This means that, for a range of decision thresholds (operating points), decisions made using \textit{T45} scores result in the rejection or acceptance of every input. It is only within a small range of thresholds that the decision cost is lower. This indicates that \textit{T45} outputs are not well calibrated.
In comparison, results shown in Fig.~\ref{fig:ana_ape}\subref{fig:ana_ape_2} indicate that \textit{T27} obtains lower decision costs across the same range of thresholds showing that system T27 is better calibrated.

As Fig.~\ref{fig:ana_score}\subref{fig:ana_scores_1} indicates, \textit{T45} produces scores in the range of 0 to 1 (likely posterior probabilities), which is incompatible with Bayes' decisions made using LLRs.
In contrast, the \textit{T27} system uses logistic-regression-based score calibration~\cite{ferrer_calibration_tutorial}, and 
hence scores are more consistent with LLRs and compatible with Bayes' decisions.

In fact, miscalibrated systems can be better calibrated with only minimal effort.
The transformation of probability-like \textit{T45} scores into LLR-like scores via a logit function $\log(y/(1-y))$~\cite[Eq.(8)]{Brummer2006-application-independent}, results in dramatic improvements (dashed line in Fig.~\ref{fig:ana_ape}\subref{fig:ana_ape_1}). 
Of course, there are other more general~\cite{Brummer2006-application-independent,ferrer_calibration_tutorial} alternatives than the logit function, which can be applied only to posterior probabilities and which is used here purely for demonstrative purposes. One such method is the logistic-regression-based calibration used by \textit{T27}. 

For reference, we plot in Fig.~\ref{fig:ana_ape}\subref{fig:ana_ape_2}, the curve obtained using the oracle pool adjacent violators (PAV) 
calibration method~\cite{Brummer2006-application-independent}.  The curve for the \textit{T27} system is close to that of the oracle curve. The simple score transformation produced using the logit function also brings the \textit{T45} system closer to an oracle calibrated version showing again that a system can be better calibrated with straightforward techniques adopted from, for example, the field of speaker verification~\cite{ferrer_calibration_tutorial}. 

\textcolor{coloredit}{In addition to \textit{T45}, all other teams which obtained an actDCF of 1.0 produced probability-like scores. Even if these scores can discriminate between bona fide and spoof/deepfakes data to varying degrees, they are not useful when users have to make decisions by comparing the systems' outputs with a pre-defined threshold. As discussed above, converting the probability-like scores into LLR-like values requires minimal implementation changes. Furthermore, better calibration towards LLRs can be achieved using existing methods. We hope this discussion highlights the importance of calibration and encourages further research in this direction within the community.}

\subsection{Cross-dataset evaluation}
\label{sec:dis:cross_db_eval}
The ASVspoof~5 evaluation set contains attacks that are generated with techniques different to those used in generating the training and development data (\S~\ref{sec:results}).
Nonetheless, with the pursuit of generalizable solutions being core to the ASVspoof initiative from its inception, we were interested to observe how well the top submissions perform when tested using data from different domains and databases.

We invited authors of the top-5 submissions to the Track~1 open condition to participate in a post-challenge, cross-dataset evaluation.
Four accepted. Using their challenge submission systems, they scored additional subsets of 3k bona fide and 3k spoof/deepfake utterances contained in the 2015, 2019 (logical access) and the 2021 (logical access and deepfake) ASVspoof challenge datasets as well as the In-the-wild (ITW) dataset~\cite{muller22_interspeech}. 
The previous ASVspoof datasets are sourced from the VCTK database~\cite{yamagishiCSTRVCTKCorpus2019}, while the ITW dataset contains bona fide and spoof/deepfake utterances of 58 celebrities and politicians, all collected from social networks and video streaming platforms. \textcolor{coloredit}{As a sanity check, we included a subset of the ASVspoof~5 Track~1 evaluation data containing 3k bona fide and 3k spoof/deepfake utterances so that we could check the correspondence to the scores in each team's initial submission.}

Results are presented in Table~\ref{tab:submissions-cross-db-results}. 
For all four systems, EERs for the smaller ASVspoof~5 Track~1 subset are similar to corresponding results for the full set shown in Figure~\ref{fig:results_all_t1}\subref{fig:result_all_t1_open}. 
However, when tested with the other ASVspoof and ITW subsets, and with only one exception (\textit{T43}, ITW), EERs increase to over 10\% for all four systems. Across the six subsets, none of the four systems performs substantially better than others.

In extending the cross-dataset evaluation, we trained a wav2vec-LLGF CM~\cite{ssl_frontends_asvspoof2021} using different combinations of ASVspoof~2015, 2019 and ASVspoof~5 training sets and evaluated detection performance using a larger set of alternative databases.
Table~\ref{tab:self-cross-db-results} shows considerable variation in performance, consistent with the findings above; while some EERs are low, others are substantially higher, and pooled and average EERs (last two rows of Table~\ref{tab:self-cross-db-results}) remain high.
The mixing of training data from different sources leads to some improvements in the EER (last four columns of Table~\ref{tab:self-cross-db-results}, especially when ASVspoof~5 and ASVspoof~2015 training data are combined.  
The best results, illustrated in bold, are all derived when the ASVspoof~5 training data is used.
Nonetheless, EERs remain high and none of the training configurations leads to acceptable EERs for the full set of databases.
Generalisation remains a challenge.

\begin{table}[t!]
    \centering
    \caption{Equal Error Rate (EER, \%) on the prepared post-evaluation package for cross-database evaluation. The four systems are among the top-5 submissions to Track 1 open condition. \textcolor{coloredit}{Each subset in the evaluation package contains a random selection of 3k bona fide and 3k spoof/deepfake utterances.}}
    \begin{tabular}{lcccc}
        \toprule       
        Evaluation subset  & \textit{T36} & \textit{T27} & \textit{T23} & \textit{T43} \\
        \midrule
        ASVspoof 5 Track1 & 3.37 & \textbf{3.30} & 4.23 & 4.33 \\
        \midrule
        ASVspoof 2015  & 10.8 & \textbf{10.40} & 12.3 & 10.6 \\
        ASVspoof 2019 LA  & \textbf{16.27} & 17.33 & 16.73 & 26.63 \\
        ASVspoof 2021 LA  & 15.73 & 18.7 & \textbf{13.13} & 25.57 \\
        ASVspoof 2021 DF  & 11.57 & \textbf{10.63} & 14.87 & 14.2 \\        
        In-the-wild   & 14.71 & 13.37 & 10.2 & \textbf{6.85} \\
        \bottomrule
    \end{tabular}
    \label{tab:submissions-cross-db-results}
\end{table}

\begin{table}[t!]
    \centering
    \caption{Equal Error Rate (EER, \%) of a wav2vec2-LLGF system trained on different permutations of the ASVspoof training sets and evaluated on different test sets. \textcolor{coloredit}{Higher EER values are indicated by darker shading, while the lowest EER value in each row is highlighted in blue.}}
    \setlength{\tabcolsep}{2pt}{
    \begin{tabular}{l|rrr|rrr|r}
        \toprule
        Trained on 2015   &  $\checkmark$  &             &            & $\checkmark$   & $\checkmark$   &             & $\checkmark$ \\
        Trained on 2019   &              & $\checkmark$  &            & $\checkmark$   &              & $\checkmark$  & $\checkmark$ \\
        Trained on 5   &              &             & $\checkmark$ &              & $\checkmark$      & $\checkmark$  & $\checkmark$ \\
        \midrule
           In the wild     & \cellcolor[rgb]{0.93, 0.93, 0.93} 12.30 & \cellcolor[rgb]{0.94, 0.94, 0.94} 10.68 & \cellcolor[rgb]{0.99, 0.99, 0.99} 2.50 & \cellcolor[rgb]{0.94, 0.94, 0.94} 10.93 & \cellcolor[rgb]{0.99, 0.99, 0.99} \textcolor{blue}{2.01} & \cellcolor[rgb]{0.99, 0.99, 0.99} 2.54 & \cellcolor[rgb]{0.99, 0.99, 0.99} 3.06\\ 
          ASVspoof 2019    & \cellcolor[rgb]{0.93, 0.93, 0.93} 11.74 & \cellcolor[rgb]{0.97, 0.97, 0.97} 6.35 & \cellcolor[rgb]{0.96, 0.96, 0.96} 8.13 & \cellcolor[rgb]{0.98, 0.98, 0.98} 5.11 & \cellcolor[rgb]{0.95, 0.95, 0.95} 8.83 & \cellcolor[rgb]{0.97, 0.97, 0.97} 5.54 & \cellcolor[rgb]{0.98, 0.98, 0.98} \textcolor{blue}{3.89}\\ 
         ASVspoof 2021 LA  & \cellcolor[rgb]{0.88, 0.88, 0.88} 17.60 & \cellcolor[rgb]{0.95, 0.95, 0.95} 8.86 & \cellcolor[rgb]{0.94, 0.94, 0.94} 10.21 & \cellcolor[rgb]{0.95, 0.95, 0.95} 9.01 & \cellcolor[rgb]{0.94, 0.94, 0.94} 10.55 & \cellcolor[rgb]{0.96, 0.96, 0.96} 8.29 & \cellcolor[rgb]{0.96, 0.96, 0.96} \textcolor{blue}{7.28}\\ 
         ASVspoof 2021 DF  & \cellcolor[rgb]{0.95, 0.95, 0.95} 9.09 & \cellcolor[rgb]{0.98, 0.98, 0.98} 4.58 & \cellcolor[rgb]{0.98, 0.98, 0.98} 5.20 & \cellcolor[rgb]{0.98, 0.98, 0.98} 4.18 & \cellcolor[rgb]{0.99, 0.99, 0.99} 3.42 & \cellcolor[rgb]{0.99, 0.99, 0.99} 2.45 & \cellcolor[rgb]{1.00, 1.00, 1.00} \textcolor{blue}{1.80}\\ 
        ASVspoof 5 Track 1 & \cellcolor[rgb]{0.86, 0.86, 0.86} 19.60 & \cellcolor[rgb]{0.94, 0.94, 0.94} 10.86 & \cellcolor[rgb]{0.94, 0.94, 0.94} 10.55 & \cellcolor[rgb]{0.92, 0.92, 0.92} 13.51 & \cellcolor[rgb]{0.93, 0.93, 0.93} 12.18 & \cellcolor[rgb]{0.95, 0.95, 0.95} \textcolor{blue}{9.06} & \cellcolor[rgb]{0.93, 0.93, 0.93} 11.67\\ 
            FakeOrReal     & \cellcolor[rgb]{0.97, 0.97, 0.97} 5.92 & \cellcolor[rgb]{0.93, 0.93, 0.93} 11.88 & \cellcolor[rgb]{0.92, 0.92, 0.92} 12.63 & \cellcolor[rgb]{0.95, 0.95, 0.95} 8.79 & \cellcolor[rgb]{0.98, 0.98, 0.98} \textcolor{blue}{5.04} & \cellcolor[rgb]{0.96, 0.96, 0.96} 7.60 & \cellcolor[rgb]{0.96, 0.96, 0.96} 8.61\\ 
            Codecfake      & \cellcolor[rgb]{0.64, 0.64, 0.64} 36.53 & \cellcolor[rgb]{0.68, 0.68, 0.68} 34.10 & \cellcolor[rgb]{0.84, 0.84, 0.84} \textcolor{blue}{21.68} & \cellcolor[rgb]{0.66, 0.66, 0.66} 35.33 & \cellcolor[rgb]{0.79, 0.79, 0.79} 25.88 & \cellcolor[rgb]{0.81, 0.81, 0.81} 24.57 & \cellcolor[rgb]{0.80, 0.80, 0.80} 25.09\\ 
            ADD2022 T1     & \cellcolor[rgb]{0.72, 0.72, 0.72} 31.46 & \cellcolor[rgb]{0.68, 0.68, 0.68} 33.90 & \cellcolor[rgb]{0.81, 0.81, 0.81} \textcolor{blue}{24.13} & \cellcolor[rgb]{0.68, 0.68, 0.68} 33.86 & \cellcolor[rgb]{0.80, 0.80, 0.80} 25.17 & \cellcolor[rgb]{0.78, 0.78, 0.78} 26.98 & \cellcolor[rgb]{0.79, 0.79, 0.79} 26.07\\ 
           ADD2022 T3.2    & \cellcolor[rgb]{0.88, 0.88, 0.88} 17.54 & \cellcolor[rgb]{0.92, 0.92, 0.92} 13.52 & \cellcolor[rgb]{0.97, 0.97, 0.97} 6.81 & \cellcolor[rgb]{0.92, 0.92, 0.92} 13.65 & \cellcolor[rgb]{0.96, 0.96, 0.96} 7.17 & \cellcolor[rgb]{0.97, 0.97, 0.97} 6.63 & \cellcolor[rgb]{0.97, 0.97, 0.97} \textcolor{blue}{5.92}\\ 
         ADD2023 T1.2 R1   & \cellcolor[rgb]{0.59, 0.59, 0.59} 39.73 & \cellcolor[rgb]{0.80, 0.80, 0.80} 25.27 & \cellcolor[rgb]{0.91, 0.91, 0.91} 14.40 & \cellcolor[rgb]{0.80, 0.80, 0.80} 25.09 & \cellcolor[rgb]{0.91, 0.91, 0.91} 14.66 & \cellcolor[rgb]{0.92, 0.92, 0.92} \textcolor{blue}{13.70} & \cellcolor[rgb]{0.88, 0.88, 0.88} 16.91\\ 
         ADD2023 T1.2 R2   & \cellcolor[rgb]{0.63, 0.63, 0.63} 37.11 & \cellcolor[rgb]{0.80, 0.80, 0.80} 25.45 & \cellcolor[rgb]{0.86, 0.86, 0.86} \textcolor{blue}{19.13} & \cellcolor[rgb]{0.81, 0.81, 0.81} 24.60 & \cellcolor[rgb]{0.86, 0.86, 0.86} 19.68 & \cellcolor[rgb]{0.86, 0.86, 0.86} 19.32 & \cellcolor[rgb]{0.84, 0.84, 0.84} 21.21\\ 
              DFADD        & \cellcolor[rgb]{0.85, 0.85, 0.85} 20.95 & \cellcolor[rgb]{0.89, 0.89, 0.89} 15.92 & \cellcolor[rgb]{1.00, 1.00, 1.00} \textcolor{blue}{1.46} & \cellcolor[rgb]{0.91, 0.91, 0.91} 14.32 & \cellcolor[rgb]{0.99, 0.99, 0.99} 2.79 & \cellcolor[rgb]{0.96, 0.96, 0.96} 7.29 & \cellcolor[rgb]{0.97, 0.97, 0.97} 5.44\\ 
            LibriSeVoc     & \cellcolor[rgb]{0.97, 0.97, 0.97} 5.68 & \cellcolor[rgb]{0.98, 0.98, 0.98} 4.17 & \cellcolor[rgb]{1.00, 1.00, 1.00} 1.97 & \cellcolor[rgb]{0.99, 0.99, 0.99} 3.60 & \cellcolor[rgb]{1.00, 1.00, 1.00} 1.55 & \cellcolor[rgb]{1.00, 1.00, 1.00} \textcolor{blue}{1.10} & \cellcolor[rgb]{1.00, 1.00, 1.00} 1.74\\ 
              Sonar        & \cellcolor[rgb]{0.86, 0.86, 0.86} 19.17 & \cellcolor[rgb]{0.70, 0.70, 0.70} 33.03 & \cellcolor[rgb]{0.79, 0.79, 0.79} 25.59 & \cellcolor[rgb]{0.62, 0.62, 0.62} 37.47 & \cellcolor[rgb]{0.90, 0.90, 0.90} \textcolor{blue}{15.48} & \cellcolor[rgb]{0.78, 0.78, 0.78} 26.25 & \cellcolor[rgb]{0.79, 0.79, 0.79} 25.64\\ 
        \midrule
              Pooled       & \cellcolor[rgb]{0.79, 0.79, 0.79} 25.95 & \cellcolor[rgb]{0.84, 0.84, 0.84} 21.84 & \cellcolor[rgb]{0.90, 0.90, 0.90} 15.20 & \cellcolor[rgb]{0.82, 0.82, 0.82} 23.15 & \cellcolor[rgb]{0.92, 0.92, 0.92} \textcolor{blue}{12.65} & \cellcolor[rgb]{0.91, 0.91, 0.91} 13.85 & \cellcolor[rgb]{0.91, 0.91, 0.91} 14.09\\ 
             Average       & \cellcolor[rgb]{0.85, 0.85, 0.85} 20.32 & \cellcolor[rgb]{0.88, 0.88, 0.88} 17.04 & \cellcolor[rgb]{0.93, 0.93, 0.93} 11.74 & \cellcolor[rgb]{0.88, 0.88, 0.88} 17.10 & \cellcolor[rgb]{0.94, 0.94, 0.94} \textcolor{blue}{11.03} & \cellcolor[rgb]{0.94, 0.94, 0.94} 11.52 & \cellcolor[rgb]{0.93, 0.93, 0.93} 11.74\\ 
        \bottomrule
    \end{tabular}
    }
    \label{tab:self-cross-db-results}
\end{table}

\subsection{Post Challenge and Related Work}
\label{sec:postchallenge}

While each ASVspoof challenge edition was designed to tackle specific research problems, post-challenge studies often uncover new directions or propose new solutions, a selection of which is discussed below.

\subsubsection{Use of foundation models}
Many submissions to the open conditions relied on the use of pre-trained foundation models. Follow-up, post-challenge studies have since explored adaptation of foundation models to the speech spoofing detection task with lower computation costs. One such study explored the projection of high-dimensional, latent features produced by a foundation model into a lower-dimensional space before classification~\cite{liu_nes2net_2025}. The use of a Res2Net-like backend, which is considerably more compact than the AASIST backend used by many challenge participants, was found to produce comparable detection performance. Other studies~\cite{pan_molex_2025,laakkonen2025mixture} investigated the use of low-rank adapters within the foundation model. Fine-tuning is then applied to the adapters instead of the entire model. 

\subsubsection{Generalization to multilingual and in-the-wild data}
ASVspoof challenges have focused exclusively on English. A notable effort in research for other languages is the Multi-Language Audio Anti-Spoof Dataset (MLAAD)~\cite{muller_mlaad_2024}, initially released during the preparation of ASVspoof~5. It paves the way to analyse detection performance in language-mismatched conditions, for example, training using ASVspoof~5 but evaluation using non-English data~\cite{moreno_revealing_2025}.  
The detection of spoof/deepfakes in unseen languages may degrade even if the system is well-trained using English data. One way to mitigate the degradation in language-mismatched conditions is to augment English-only training data with accented English data generated by text-to-speech synthesis systems~\cite{10832142}.

\subsubsection{Data-Centric Approach} 

Recent work~\cite{combei2025unmasking} has investigated data-centric approaches to reduce redundancy, label-noise, and speaker/gender imbalances that can undermine model robustness and generalisation. Performance can be improved by training using dataset pruning strategies~\cite{combei2025unmasking}, such as diversity-aware subset selection via (i)~data-informed pruning, which keeps either the most representative (closest to a class prototype) or the most diverse (furthest from the class mean) samples based on the embedding distance, and (ii)~algorithm-informed pruning, which removes unreliable samples near the decision boundary and extreme outliers using logistic-regression margins. These pruning techniques are shown to match or exceed performance for full-dataset training, while also improving generalisation to unseen spoofing attacks.

\subsection{Limitations and Future Directions}
\label{sec:future}
As with every challenge and benchmarking exercise, it is important to acknowledge and understand the limitations. A selection of the most pertinent limitations and other issues raised by participants are discussed in the following.

\subsubsection{Beyond binary classification}
Speech spoofing detection is framed as a binary classification task.  There is also a developing interest in multi-class source tracing or attribution~\cite{yan2022initial,zhusource,kleinsource,MISHRA2026101840} for which the aim is to identify or characterize the particular approach, algorithm, tool or model/architecture components used in the generation of spoofed data. 
Source tracing can be used to help link different spoof/deepfake data produced by a common source, for accountability, and hence to encourage generative speech technology creators, services or platforms to harden tools against misuse.\footnote{\textcolor{coloredit}{An alternative, proactive strategy involves the embedding of imperceptible watermarks~\cite{roman2024proactive}) into either bona fide or spoof/deepfake utterances. The bona fide / spoof detection task is then recast as a watermark detection task. While this and other similar strategies are beyond the scope of passive spoofing detection methods analysed in this paper, these approaches can serve as complementary defence layers.}}

Recent studies presented at the Interspeech 2025 special session on \textit{Source Tracing: The Origins of Synthetic or Manipulated Speech} include open-set multi-class classification techniques to characterise previously unseen spoof/deepfake attacks~\cite{klein2025open,stan2025tada}, neural codec class tracing~\cite{xie2025neural,chen2025codec,chen2025towards},
a source verification task that tests whether two spoofed samples were produced using the same generator~\cite{koutsianos2025synthetic,kulkarni2025unveiling, firc2025stopa,falez2025audio}, and explainable source tracing~\cite{yunbin2025sourcetracing}.

\subsubsection{Definition of spoofed speech}
One of the questions raised by some participants focuses on the very definition of a spoofed speech sample. The potential ambiguity stems primarily from the use of neural audio codecs in ASVspoof~5.
Neural codecs can introduce artefacts that are similar to those introduced using vocoders commonly employed in TTS and VC techniques. Consequently, bona fide speech processed using a neural codec, may exhibit artefacts that resemble those embedded in spoofs/deepfakes. 

While for ASVspoof~5, spoofs/deepfakes are defined based on their generation using TTS or VC, it is clear that the detection of mere vocoding artefacts may no longer serve as a reliable indicator. The distinction between bona fide and spoofed speech is thus arguably narrowing. Furthermore, other operations such as neural speech enhancement might also introduce artefacts into bona fide speech that resemble those in spoofs/deepfakes. 
Therefore, the definition of what constitutes a spoof, much like the artefacts used to distinguish AI-generated from real speech, should evolve and requires discussion and reflection in the future.

\subsubsection{Source data diversity}

The acquisition and reliance on a single corpus (e.g., VCTK or MLS/LibriVox) for constructing bona fide speech samples has been a recurring criticism in the community. Such data does not reflect the variability seen in the wild where recording conditions, devices, and speaker populations vary much more widely~\cite{kwok_bona_2025}. 
While progress has been made in this respect, by using data for ASVspoof~5 collected in more diverse recording setups (different rooms, microphones, and speakers), the scenario remains somewhat narrow, focusing on audiobook-style read speech. The resulting data variability may thus still be far from the heterogeneity of speech encountered in the wild.

On the other hand, it remains important to recognise the value of carefully controlled \emph{training} conditions. 
When bona fide material is homogeneous, the discriminative cues learned by detection models are more likely to arise from spoofing artifacts rather than from incidental differences in domains, channels, or recording environments. However, evaluation data could, and arguably should, include bona fide and spoofed speech drawn from different domains and scenarios to better assess generalisation.

\subsubsection{Algorithmic innovation of modern speech spoofing detectors}
The analysis of top-performing systems summarized in Table~\ref{tab:system_summary}, across both tracks and conditions, reveals a problem of concern: while data augmentation and score/system fusion strategies vary widely between top submissions, core model architectures, specifically \emph{acoustic frontends} and \emph{backend classifiers}, are becoming homogeneous. 
\textcolor{coloredit}{For the closed condition, the combination of Mel spectrograms with ResNet (T47 and T24, both Tracks) and waveform inputs with RawNet (T24 and T45, both tracks) emerge as popular CM implementations. 
Both combinations were also employed by top submissions to the ASVspoof 2021 challenge~\cite[Figs. 9 and 11]{liuASVspoof2021Spoofed2023}.
For the open condition, wav2vec2 and WavLM models are utilised by all the top submissions. Their combination with existing backends is similar to methods published well before the ASVspoof~5 challenge~\cite{donasVicomtech2022,ssl_frontends_asvspoof2021,wav2vec_antispoofing2022}.}

Such observations suggest that architectural innovations in speech spoofing detection may be reaching a bottleneck.
Meanwhile, ongoing progress in the detection of spoofed speech artifacts is heavily dependent on extrinsic factors such as principled data design, adaptive fusion strategies, and a deeper understanding of generalization across conditions. These issues demand greater attention in the future to address architectural homogeneity and to explore alternative model paradigms beyond those based on SSL frontends and well-established binary classifiers.

\subsubsection{Generalisation to diverse attacks}

A closer inspection of Figure~\ref{fig:ana_t1}\subref{fig:ana_selected_attacks_t1_closed} (and 
Figure 7(a)
in the appendix), which displays closed condition results for the top-3 systems, reveals clear variability in system behaviour across different attacks. The distinct markers representing individual systems indicate that no single approach consistently dominates across all attack types. In several cases (e.g., A18 vs. A21), the relative ranking of systems is inverted.  

This pattern suggests a limited ability of models to generalise beyond the specific spoofing characteristics encountered during training, reflecting a degree of attack-dependent overfitting. Such behaviour implies that systems have learned decision boundaries that are highly tuned to the acoustic or generative traits of specific spoofing families rather than capturing more robust, attack-invariant cues. The large range in minDCF values across attacks further supports this interpretation, as systems that achieve near-optimal performance on some attacks can degrade severely on others, including the legacy A19 unit selection attack. Overall, results highlight the challenge of building generalised countermeasures capable of generalising to diverse spoofing attacks with closed data constraints. 

\textcolor{coloredit}{Finally, in certain application scenarios, spoof/deepfake attacks can be replayed, re-recorded, or convolved with room reverberation before presentation.
Such processed data were found to be particularly challenging to detect~\cite{kirill2025safe,muller_replay_2025}. 
This scenario extends beyond the replaying of bona fide recordings, i.e., the physical access scenario in previous ASVspoof challenge editions. The compounding of room acoustics with spoof/deepfake artifacts represents a key challenge to detection and remains within scope for
future research and challenge editions.
}

\section{Conclusions}
\label{sec:conclusions}
The ASVspoof initiative and challenge series are designed to foster progress in spoof/deepfake speech detection and spoofing-robust automatic speaker verification (SASV).  
As for all previous editions, ASVspoof~5 brings several advances and new challenges. 
\textcolor{coloredit}{
It incorporates adversarial attacks, bona fide and spoofed data collected or generated from a substantially larger speaker population recorded under variable recording conditions. Spoofs/deepfakes were generated with the very latest (as of 2024) generative speech technology and treated with contemporary encoding/compression techniques, while a new open condition with a relaxed training data policy was adopted for the first time.
}
With a full description of the database being available elsewhere, in this paper we provide an overview of the ASVspoof~5 challenge results and analyses. 
We also report new analyses of score calibration and cross-dataset evaluation using top submissions. 
Results show promising detection performance, but also reveal some limitations of both the challenge and detection solutions.

Results indicate a persistent lack of generalization to spoofed data generated using different attack techniques, particularly under closed training conditions in which the data used for training is restricted.
While the use of foundation models under open training conditions leads to substantially more reliable detection performance, the cross-dataset evaluation shows that even the best performing systems, as judged from evaluation using ASVspoof~5 data, yield notably higher detection error rates when evaluation is performed using out-of-domain evaluation datasets as well as previous ASVspoof challenge databases. Current detection solutions overfit to the acoustic or generative traits of specific datasets.
Generalization remains a holy grail in speech spoof/deepfake detection.
With many of the top performing systems using homogenous model architectures,
breakthroughs may come from the continued exploration of novel model architectures, but may also come from more principled data design, better fusion strategies, data augmentation techniques, and model training paradigms beyond supervised training.

Future editions of ASVspoof must hence continue the search for better-generalisable detection solutions. 
More diverse source data in terms of languages, speakers, and recording conditions must also be considered.
With ASVspoof~5 having also exposed calibration issues in spoof/deepfake detection, and in mirroring trends in the evaluation of automatic speaker verification systems, calibration-sensitive metrics may be adopted as primary evaluation metrics in future editions.

\section*{Acknowledgments}
The ASVspoof 5 organising committee extends its sincere gratitude to challenge participants (anonymised) and data contributors listed in the database paper~\cite{wang_asvspoof5database}.
This work is partially supported by JST, PRESTO Grant Number JPMJPR23P9, Japan, and with funding received from the French Agence Nationale de la Recherche (ANR) via
the BRUEL (ANR-22-CE39-0009) and COMPROMIS (ANR22-PECY-0011) projects. This work was also partially supported by the Academy of Finland (Decision No. 349605, project ``SPEECHFAKES''), and the Innovation and Technology Fund, Hong Kong SAR (MHP/048/24). Part of the computation and data generation is carried out using the TSUBAME4.0 supercomputer at Institute of Science Tokyo (Japan).

\section*{References}
{
\printbibliography
}

\clearpage
\newpage
\appendix
\subsection{\textcolor{coloredit}{Note on actDCF and minDCF}}

Here we show that the value minDCF is equal to that of actDCF when CM is perfectly calibrated. This is already mentioned in other studies (e.g., [51](\S~2.5.2)). Let us recap the definition of DCF in Eq.~(\ref{eq:dcf-track1-norm}), which is
\begin{equation}
\text{DCF}(\tau_\text{cm}) = \beta \cdot P_\text{miss}^\text{cm}(\tau_\text{cm}) + P_\text{fa}^\text{cm}(\tau_\text{cm}).
\end{equation}
Let $p({{S}}|C_\text{Bon})$ and $p({{S}}|C_\text{Spf})$ denote the probability density function (PDF) of the score $R$ for the classes of bona fide and spoof, respectively. Note that we differentiate the random variable $S$ from its value $s$. 
We can then write the miss rate $P_\text{miss}^\text{cm}(\tau_\text{cm})$ and false alarm rate $P_\text{fa}^\text{cm}(\tau_\text{cm})$ as
\begin{equation}
    P_\text{miss}^\text{cm}(\tau_\text{cm}) = \int_{-\infty}^{\tau_\text{cm}} p(S=s|C_\text{Bon}) ds
\end{equation}
and
\begin{equation}
    P_\text{fa}^\text{cm}(\tau_\text{cm}) = \int_{\tau_\text{cm}}^{+\infty} p(S=s|C_\text{Spf}) ds,
\end{equation}
respectively. Note that $P_\text{miss}^\text{cm}(\tau_\text{cm})$ is a monotonic increasing function of $\tau_\text{cm}$ while $P_\text{fa}^\text{cm}(\tau_\text{cm})$ is a monotonic decreasing function of $\tau_\text{cm}$.

Let us now find the decision threshold $\tau_\text{cm}^*$ for the minDCF. By taking the derivative of DCF$(\tau_\text{cm})$ w.r.t. $\tau_\text{cm}$, we get
\begin{equation}
    \frac{d \text{DCF}(\tau_\text{cm})}{d\tau_\text{cm}} = \beta \cdot p(S=\tau_\text{cm}|C_\text{Bon}) - p(S=\tau_\text{cm}|C_\text{Spf}).
\end{equation}
By setting $\frac{d \text{DCF}(\tau_\text{cm})}{d\tau_\text{cm}}=0$, we get the solution $\tau_\text{cm}^*$ that satisfies
\begin{equation}
    \frac{p(S=\tau_\text{cm}^*|C_\text{Bon})}{p(S=\tau_\text{cm}^*|C_\text{Spf})}
    = \frac{1}{\beta},
\end{equation}
or, equivalently, 
\begin{equation}\label{eq:app:tau_1}
    \log \frac{p(S=\tau_\text{cm}^*|C_\text{Bon})}{p(S=\tau_\text{cm}^*|C_\text{Spf})}
    = -\log{\beta}.
\end{equation}
If DCF$(\tau_\text{cm}^*)$ is the global minimum, it corresponds to the minDCF.\footnote{For example, if $p(S|C_\text{Bon})$ and $p(S|C_\text{Spf})$ are Gaussian distributions with the same variance and the mean of $p(S|C_\text{Bon})$ is larger than that of $p(S|C_\text{Spf})$, it can be shown that the second-order derivative $\frac{d^2 \text{DCF}(\tau_\text{cm})}{d\tau_\text{cm}^2} > 0$, and DCF$(\tau_\text{cm}^*)$ corresponds to the global minimum. This is not guaranteed to be true if the two distributions have different variance values or are not unimodal distributions.} 
Notice that the left hand side of the above equation is referred to as the LLR of score $S$, i.e., $\mathrm{LLR}(s)=\log \frac{p(S=s|C_\text{Bon})}{p(S=s|C_\text{Spf})}$. If $\mathrm{LLR}(s)$ is monotonic, we can easily find the $\tau_\text{cm}^*$ by using the inverse of the LLR and setting $\tau_\text{cm}^* = \mathrm{LLR}^{-1}(-\log \beta)$. Otherwise, we have to search for $\tau_\text{cm}^*$ that achieves the minimum DCF value, which is the approach used in the ASVspoof~5 challenge. 


For a perfectly calibrated CM, the system produces an LLR value $r$ which satisfies the property that `the LLR of the LLR is LLR' (ref. [61](\S~2.4)). This can be written as  
\begin{equation}
    r = \mathrm{LLR}(r) = \log \frac{p(R=r|C_\text{Bon})}{p(R=r|C_\text{Spf})},
\end{equation}
where $R$ denotes the random variable LLR. Consequently, 
the threshold achieving the minDCF of the perfectly calibrated system can be easily obtained as
\begin{equation}\label{eq:app:tau_2}
    r^* = \log \frac{p(R=r^*|C_\text{Bon})}{p(R=r^*|C_\text{Spf})} = -\log\beta.
\end{equation}
Since the Bayes decision threshold is set to $\tau_\text{Bayes}=-\log\beta$ by definition (\S~\ref{sec:metrics}), we get $r^* = \tau_\text{Bayes}$. The decision threshold achieving minDCF is equal to the Bayes decision threshold, and the values actDCF and minDCF are equal. 

Comparison between Eqs.(\ref{eq:app:tau_1}) and (\ref{eq:app:tau_2}) helps to interpret what calibration means. Since $\beta$ is decided by the requirements of the application (i.e., decision costs and priors), users of the CM have to set $\tau_\text{cm}^*$ according to $\beta$. For a not-well calibrated system, we either do $\tau_\text{cm}^* = \mathrm{LLR}^{-1}(-\log \beta)$ or search $\tau_\text{cm}^*$ over a test set. However, neither is feasible because we usually do not know $\mathrm{LLR}^{-1}$ or the test set with ground truth labels. 
Given a perfectly calibrated CM, we simply set the threshold to $r^*=-\log\beta$. 

\subsection{\textcolor{coloredit}{Summary of relevant challenges}}

{Table~\ref{tab:deepfake_challenges} summaries relevant events on speech anti-spoofing or deepfake detection challenges.}

\subsection{Full set of analysis results}
We present a full set of results analyses. 

\begin{figure*}[t!]
    \centering
    \subfloat[Attacks in closed condition]
    {\includegraphics[width=0.5\textwidth, trim=0 5 0 17, clip]    {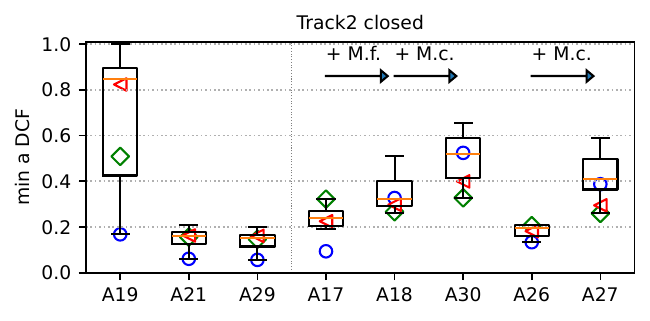}
    \label{fig:ana_selected_attacks_t2_closed}
    }
    \subfloat[Attack groups in closed (left) and open (right) conditions]
    {\includegraphics[width=0.5\textwidth, trim=0 5 0 7, clip]    {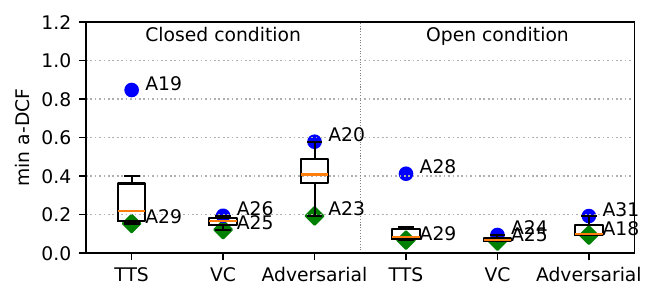}
    \label{fig:ana_merged_attacks_t2}
    }
    \hfill
    \subfloat[Codec groups in closed (left) and open (right) conditions]
    {\includegraphics[width=0.5\textwidth, trim=0 5 0 7, clip]    {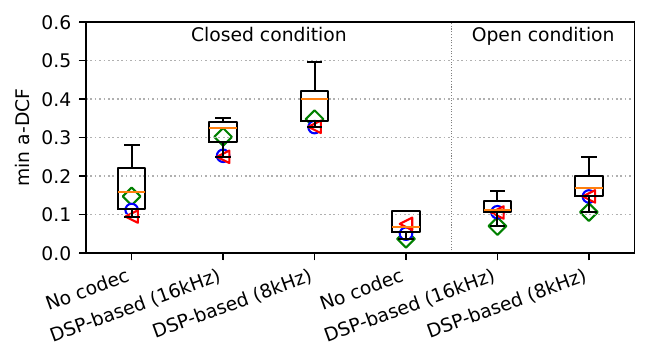}
    \label{fig:ana_merged_codec_t2}
    }
    \caption{Boxplots of evaluation set minDCF of Track 2. 
    In sub-figure (a), each box shows the raw minDCF values of top 50\% submissions in the closed condition. Markers are top-1 submission (\textcolor{teal}{$\diamond$}), top-2 (\textcolor{blue}{o}), and top-3 (\textcolor{red}{$\triangleleft$}) submissions. 
    The annotated arrows `+ M.f.' and `+ M.c.' mean that attacks are the right hand side are obtained via applying Malafide annd Malacopula, respectively, to the attacks on the left hand side. 
    Figures for other tracks and conditions are presented in the appendix.
    In sub-figure (b), the median minDCF value of the top 50\% submissions for each attack is computed, and each box summarizes the median minDCF values of the attacks in the group (either TTS, VC, or adversarial). Markers are easiest (\textcolor{teal}{$\blacklozenge$}) and hardest (\textcolor{blue}{$\bullet$}) attacks. 
    In sub-figure (c), each box shows the raw minDCF values of top 50\% submissions in a codec condition. Markers are the same as (a). 
    }
    \label{fig:ana_t2}
\end{figure*}

\begin{figure*}
    \centering
    \subfloat[Track 1 closed condition]{\includegraphics[trim=0 0 0 20, clip, width=\textwidth]{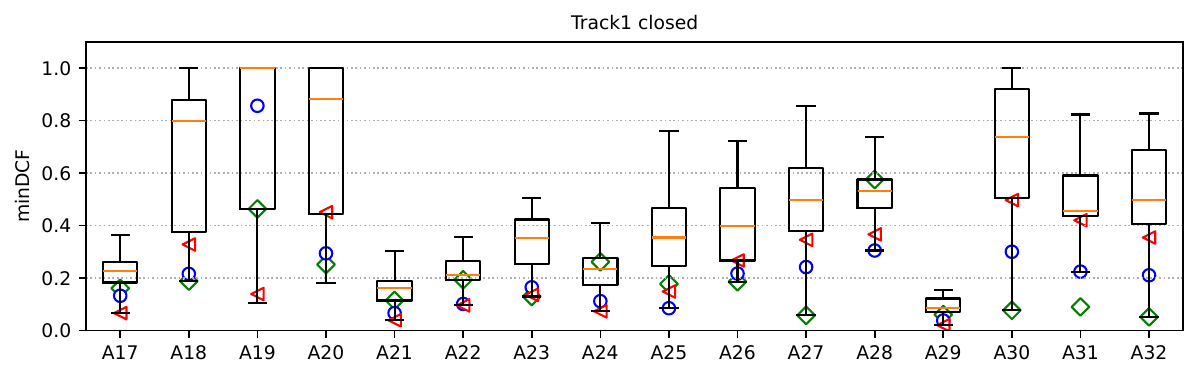}\label{fig:ana_all_attack_t1_close}}
    \hfill
    \subfloat[Track 1 open condition]{\includegraphics[trim=0 0 0 20, clip, width=\textwidth]{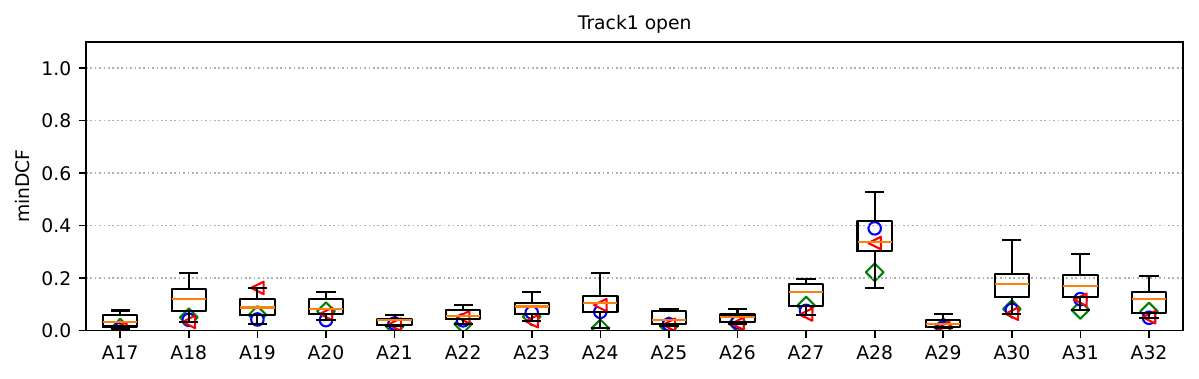}\label{fig:ana_all_attack_t1_open}}
    \hfill
    \subfloat[Track 2 closed condition]{\includegraphics[trim=0 0 0 20, clip, width=\textwidth]{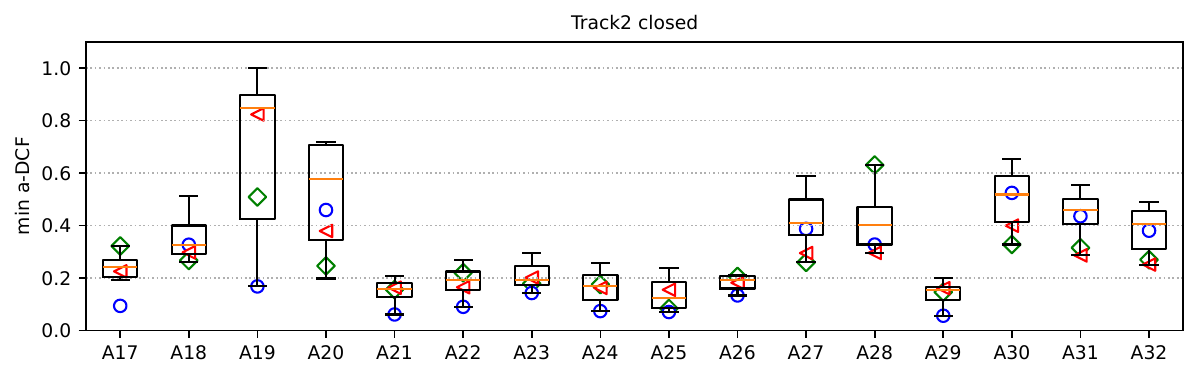}\label{fig:ana_all_attack_t2_close}}
    \hfill
    \subfloat[Track 2 open condition]{\includegraphics[trim=0 0 0 20, clip, width=\textwidth]{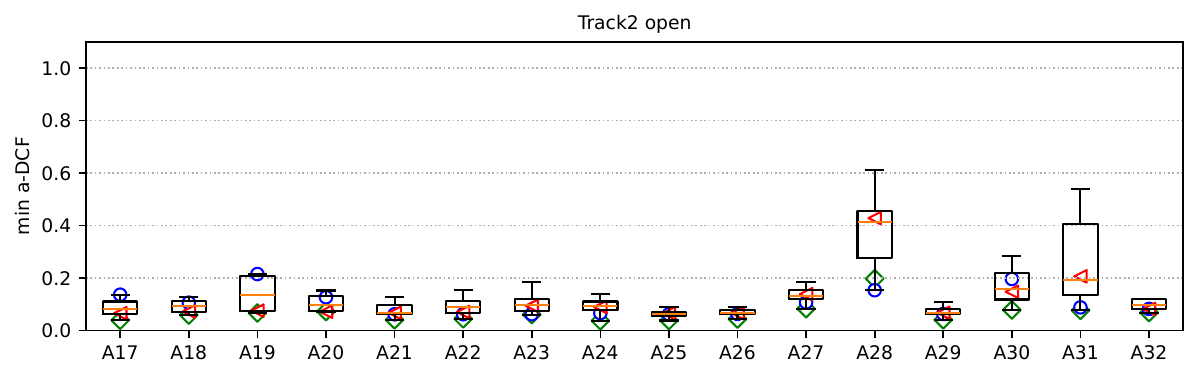}\label{fig:ana_all_attack_t2_open}}
    \caption{Boxplots of performance on detecting attacks in evaluation set. Results of the top half of  submissions are used. Markers are top-1 submission (\textcolor{teal}{$\diamond$}), top-2 (\textcolor{black}{o}), and top-3 (\textcolor{red}{$\triangleleft$}) submissions. }
    \label{fig:attackwise_track1_results}
    
\end{figure*}

\begin{figure*}
    \centering
    \subfloat[Track 1 closed condition]{\includegraphics[trim=0 0 0 15, clip, width=0.9\textwidth]{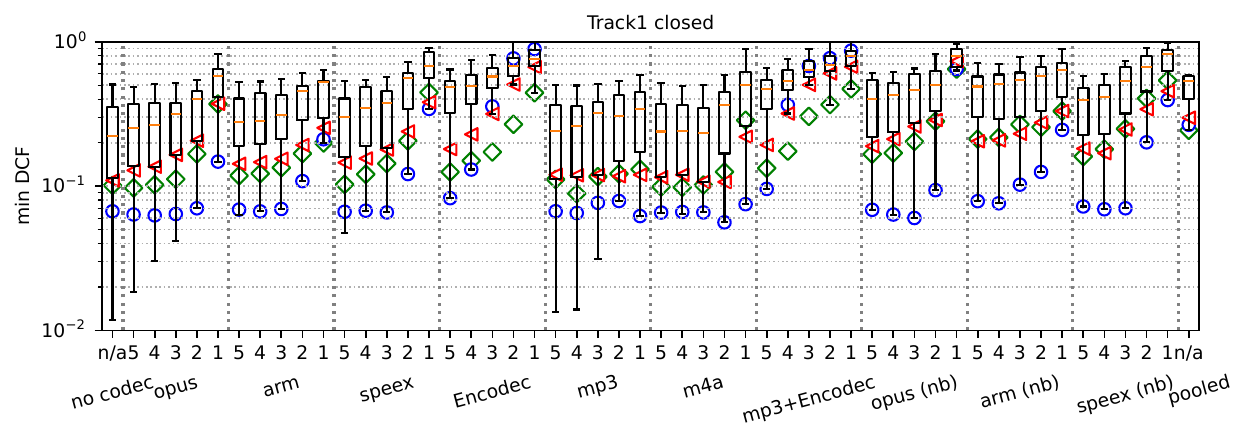}\label{fig:ana_codec_q_t1_close}}
    \hfill
    \subfloat[Track 1 open condition]{\includegraphics[trim=0 0 0 15, clip, width=0.9\textwidth]{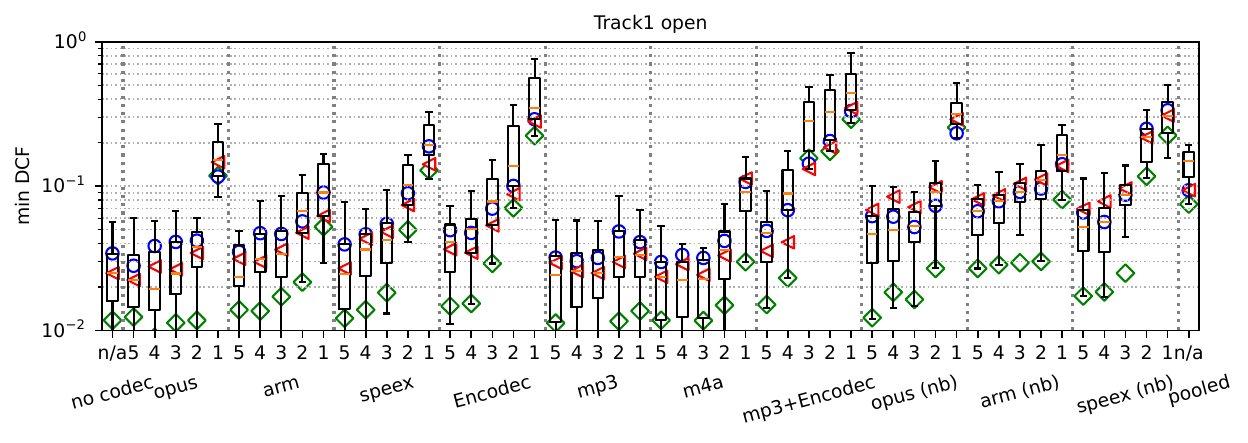}\label{fig:ana_codec_q_t1_open}}
    \hfill
    \subfloat[Track 2 closed condition]{\includegraphics[trim=0 0 0 15, clip, width=0.9\textwidth]{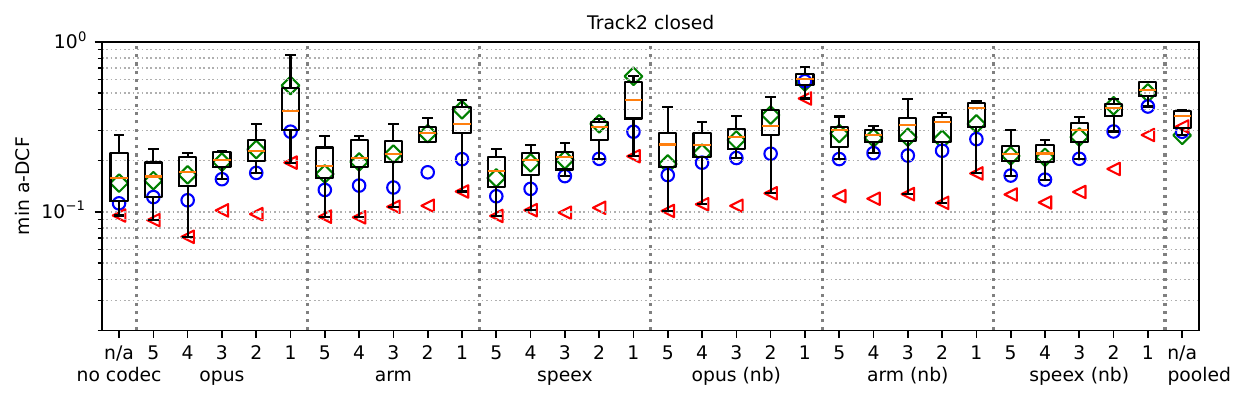}\label{fig:ana_codec_q_t2_close}}
    \hfill
    \subfloat[Track 2 open condition]{\includegraphics[trim=0 0 0 15, clip, width=0.9\textwidth]{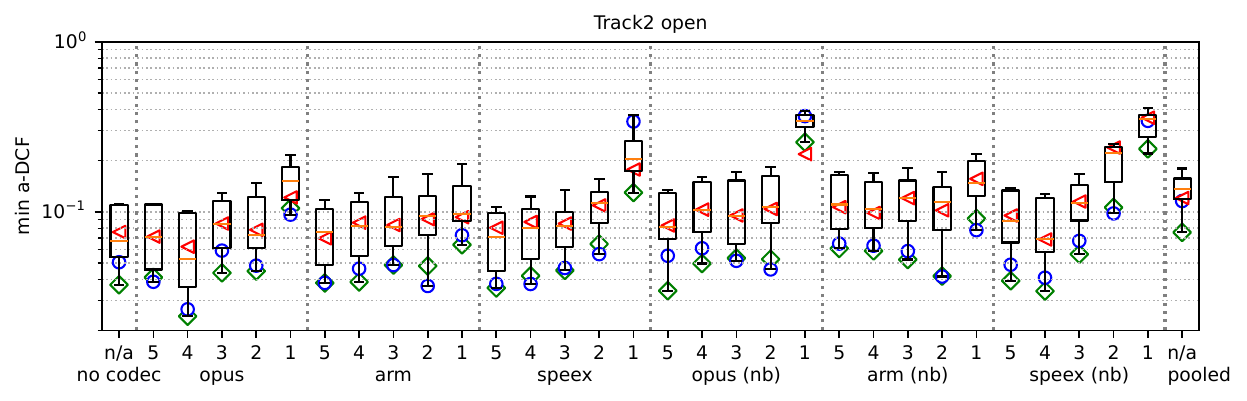}\label{fig:ana_codec_q_t1_opeen}}
    \caption{Boxplots of performance in each combination of the codecs and quality factors. Results of the top half of submissions are used. Markers are top-1 submission (\textcolor{teal}{$\diamond$}), top-2 (\textcolor{black}{o}), and top-3 (\textcolor{red}{$\triangleleft$}) submissions. }
    \label{fig:ana_codec_all}
    
\end{figure*}

\begin{figure*}
    \centering
    \subfloat[Track 1 closed condition]{\includegraphics[width=0.5\textwidth]{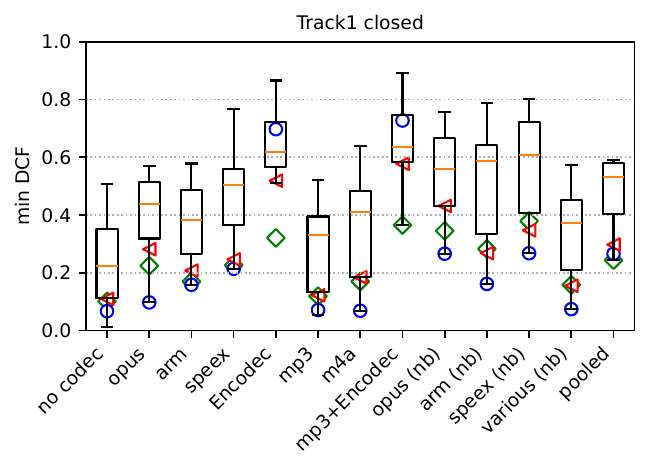}
    \label{fig:app_ana_codec_t1_closed}
    }
    \subfloat[Track 1 open condition]{\includegraphics[width=0.5\textwidth]{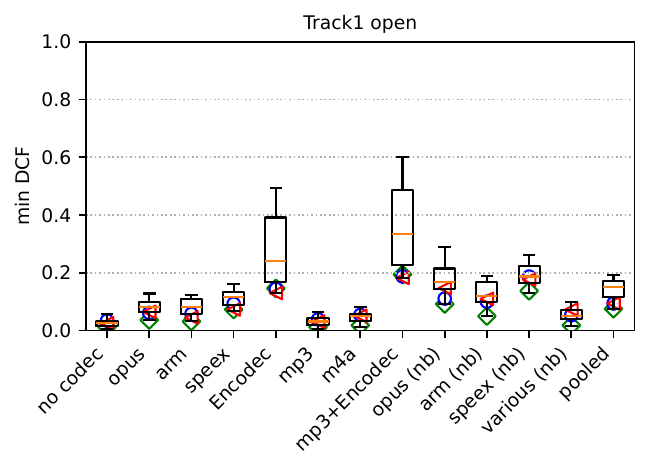}
    \label{fig:app_ana_codec_t1_open}
    }
    \hfill
    \subfloat[Track 2 closed condition]{\includegraphics[width=0.5\textwidth]{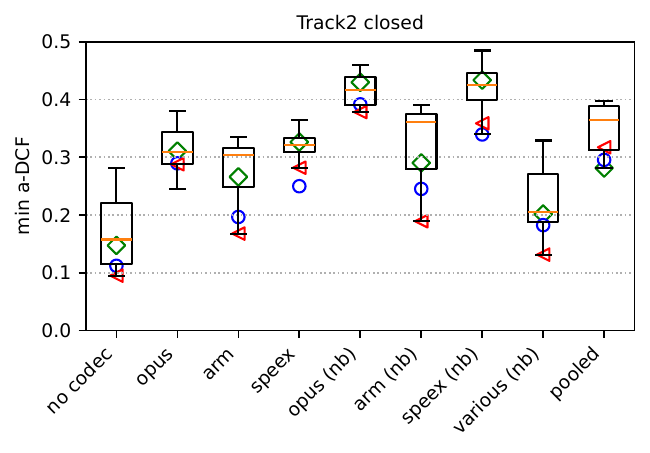}
    \label{fig:app_ana_codec_t2_closed}
    }
    \subfloat[Track 2 open condition]{\includegraphics[width=0.5\textwidth]{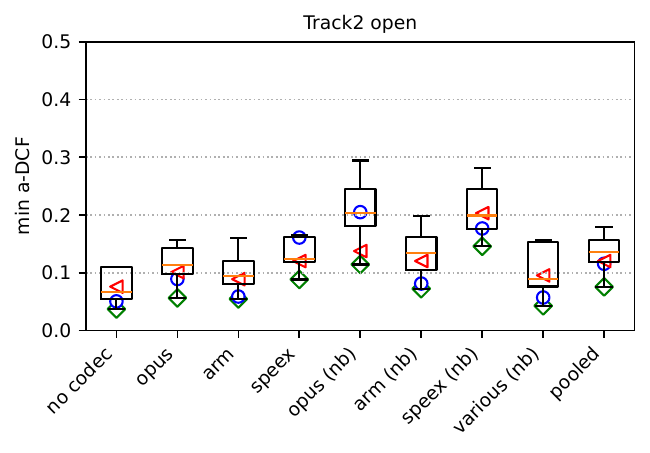}
    \label{fig:app_ana_codec_t2_open}
    }
    \caption{Boxplots of performance in different encoding conditions. Results of the top half of submissions are used. Markers are top-1 submission (\textcolor{teal}{$\diamond$}), top-2 (\textcolor{black}{o}), and top-3 (\textcolor{red}{$\triangleleft$}) submissions.
    \label{fig:codecSummary}}
\end{figure*}

\begin{itemize}
    \item Figure~\ref{fig:ana_t2} shows a visualisation of results for Track 2 and selected conditions: selected individual attacks (Figure~\ref{fig:ana_t2}\subref{fig:ana_selected_attacks_t2_closed}), a comparison between closed and open conditions (Figure~\ref{fig:ana_t2}\subref{fig:ana_merged_attacks_t2}), and the impact of codecs and compression (Figure~\ref{fig:ana_t2}\subref{fig:ana_merged_codec_t2}). The results are discussed in \S~\ref{sec:results:t2}.
    \item Figure~\ref{fig:attackwise_track1_results} shows results for primary metrics computed for each attack in the evaluation set. 
    \item Figure~\ref{fig:ana_codec_all} shows results for primary metrics computed for each combination of codec or compression condition and quality factor. The quality factor corresponds to the bit rate. The correspondence is described in Table~\ref{tab:codec_bitrates_compact}. Note that the y-axis is log-scaled.
    \item Figure~\ref{fig:codecSummary} shows pooled results of Figure~\ref{fig:ana_codec_all} over the quality factor and results for each codec and compression condition. 
\end{itemize}

\begin{table}[h!]
\centering
\caption{Bitrate levels (kbps) of codecs at levels 1--5. Abbreviate `nb' refers to the condition using an 8~kHz effective band-width.}
\begin{tabular}{lrrrrr}
\toprule
& \multicolumn{5}{c}{Codec factor ID} \\
\cmidrule{2-6}
{Codec} & {1} & {2} & {3} & {4} & {5} \\
\midrule
opus     & 6.00   & 12.00  & 18.00  & 24.00  & 30.00  \\
arm       & 6.60   & 8.85   & 14.25  & 18.25  & 23.05  \\
speex    & 5.75   & 9.80   & 16.80  & 23.80  & 34.20  \\
encodec  & 1.50   & 3.00   & 6.00   & 12.00  & 24.00  \\
mp3      & 45-85  & 80-120 & 120-150 & 170-210 & 220-260 \\
m4a      & 16.00  & 32.00  & 64.00  & 96.00  & 128.00 \\
opus (nb)     & 4.00   & 8.00   & 12.00  & 16.00  & 20.00  \\
arm (nb)      & 4.75   & 6.70   & 8.85   & 10.20  & 12.20  \\
speex (nb)    & 3.95   & 5.95   & 11.00  & 18.20  & 24.60  \\
\bottomrule
\end{tabular}
\label{tab:codec_bitrates_compact}
\end{table}

\begin{table*}[!t]
\centering
\caption{\textcolor{coloredit}{Summary of speech anti-spoofing and deepfake detection challenges}}
\label{tab:deepfake_challenges}
\small
\setlength{\tabcolsep}{2pt}{\begin{tabular}{@{}lllll@{}}
\toprule
Challenge & {Primary task} & {Data sources} & {Language} & {Defining characteristic} \\ \midrule
ASVspoof 2015 & Spoofing/fake detection & VCTK & En & 1st large-scale challenge \\
ASVspoof 2019 & Spoofing/deepfake detection & VCTK & En & DNN-based TTS/VC \\
ASVspoof 2021 LA & Spoofing/deepfake detection & VCTK & En & +Codecs \\
ASVspoof 2021 DF & Deepfake Detection & VCTK, VCC 2020 & En, De, Zh, Fi & Mix of data from VCTK and VCC \\
\textbf{ASVspoof 5} & Spoofing/deepfake detection and SASV & LibriVox & En & More diverse data, adv. attack, calibration \\ 
ADD 2022 & Spoofing/deepfake, partial spoof detection & AISHELL-3 & Zh & Partial spoof data \\
ADD 2023 & Spoofing/deepfake detection, source tracing & AISHELL-3 & Zh & Identification of vocoder \\
SAFE 2025 & Spoofing/deepfake \& manipulation detection & $>$20 sources & Multi. & Diverse data and manipulation \\
IS2025 Source Tracing & Source Tracing & MLAAD & 35+ Lang. & Cross-lingual model attribution \\ \bottomrule
\end{tabular}}
\end{table*}

\end{document}